\documentclass[11pt,a4paper,final]{article}
\usepackage{amsmath}%
\usepackage{amstext}%
\usepackage{amssymb}%
\usepackage{showkeys}%
\usepackage{epsfig}%
\usepackage{graphicx}%
\usepackage{xcolor}
\usepackage{multicol}
\usepackage{amsthm}
\usepackage{booktabs}
\usepackage{apacite}
\usepackage{multirow}
\usepackage[top=1in, bottom=1.25in, left=0.8in, right=0.8in]{geometry}

\theoremstyle{prop}
\theoremstyle{proof}

\newtheorem{proposition}{Proposition}

\providecommand{\keywords}[1]{
\textbf{Keywords:~~~} Bayesian MS--VAR process, Stochastic DDM, Gordon's Growth Model.
}

\begin{document}
\title{Bayesian Markov--Switching Vector Autoregressive Process}
\author{Battulga Gankhuu\footnote{Department of Applied Mathematics, National University of Mongolia; E-mail: battulgag@num.edu.mn; Phone Number: 976--99246036}}
\date{}

\maketitle 

\begin{abstract}
This study introduces marginal density functions of the general Bayesian Markov--Switching Vector Autoregressive (MS--VAR) process. In a special case of the Bayesian MS--VAR process, we provide closed--form density functions and Monte--Carlo simulation algorithms, including the importance sampling method. The Monte--Carlo simulation method departs from the previous simulation methods because it removes the duplication in a regime vector. To obtain smoothed probability inference, we develop a new smoothing method.
\end{abstract}



\section{Introduction}

Classic Vector Autoregressive (VAR) process was proposed by \cite{Sims80} who criticize large--scale macro--econometric models, which are designed to model interdependencies of economic variables. Besides \citeA{Sims80}, there are some other important works on multiple time series modeling, see, e.g., \citeA{Tiao81}, where a class of vector autoregressive moving average models was studied. For the VAR process, a variable in the process is modeled by its past values and the past values of other variables in the process. After the work of \citeA{Sims80}, VARs have been used for macroeconomic forecasting and policy analysis. However, if the number of variables in the system increases or the time lag is chosen high, then too many parameters need to be estimated. This will reduce the degrees of freedom of the model and entail a risk of over--parametrization. 

Therefore, to reduce the number of parameters in a high--dimensional VAR process, \citeA{Litterman79} introduced probability distributions for coefficients that are centered at the desired restrictions but that have a small and nonzero variance. Those probability distributions are known as Minnesota prior in Bayesian VAR (BVAR) literature, which is widely used in practice. Due to over--parametrization, the generally accepted result is that the forecast of the BVAR model is better than the VAR model estimated by the frequentist technique. Research works have shown that BVAR is an appropriate tool for modeling large data sets; for example, see \citeA{Banbura10}.

Sudden and dramatic changes in the financial market and economy are caused by events such as wars, market panics, or significant changes in government policies. To model those events, some authors used regime--switching models. The regime--switching model was introduced by seminal works of \citeA{Hamilton89,Hamilton90} (see also books of \citeA{Hamilton94} and \citeA{Krolzig97}), and the model is hidden Markov model with dependencies; see \citeA{Zucchini16}. However, Markov regime--switching models have been introduced before Hamilton (1989), see, \citeA{Goldfeld73}, \citeA{Quandt58}, and \citeA{Tong83}. The regime--switching model assumes that a discrete unobservable Markov process randomly switches among a finite set of regimes and that a particular parameter set defines each regime. The model fits some financial data well and has become popular in financial modeling, including equity options, bond prices, and others. 

A model that considers all of the above is the Bayesian Markov--Switching VAR (MS--VAR) process. Its applications in finance can be found in \citeA{Battulga23a}, \citeA{Battulga24f}, and \citeA{Battulga24a}. In some existing option pricing models, the underlying asset price is governed by some stochastic process, and economic variables such as GDP, inflation, unemployment rate, and so on are not taken into account. For this reason, the author has developed option pricing models, depending on economic variables. Applying the Bayesian MS--VAR process, with direct calculation and change of probability measure for some frequently used options, \citeA{Battulga24a} derived pricing formulas. Also, the author used the Bayesian MS--VAR process to price equity--linked life insurance products and rainbow options, see \citeA{Battulga23a} and \citeA{Battulga24f} .

Monte--Carlo simulation methods using the Gibbs sampling algorithm for Bayesian MS--VAR process are proposed by some authors. In particular, the Monte--Carlo simulation method of the Bayesian MS--AR($p$) process is provided by \citeA{Albert93}, and its multidimensional extension is given by \citeA{Krolzig97}. In this paper, we introduce a new Monte--Carlo simulation method that removes duplication in a regime vector. We also introduce importance sampling method to estimate probability of rare event, which corresponds to endogenous variables. Importance sampling is an effective variance reduction technique for studying the rare events. \citeA{Glasserman00} used the importance sampling method to model portfolio loss random variable by using approximation. Also, \citeA{Glasserman05b} study a loss random variable of credit portfolio  applying the method, see also \citeA{McNeil05}.

The rest of the paper is organized as follows: In Section 2, for the general Bayesian MS--VAR process, we obtain some conditional density functions, which are helpful for general Monte--Carlo simulation. Section 3 is dedicated to studying a special case of the process, where we obtain closed-form conditional density functions of our model's random components. Some of the conditional density functions have not been explored before. In Section 3, we provide Monte--Carlo simulation methods, including the importance sampling method. Finally, Section 4 concludes the study.

\section{Bayesian MS--VAR process}

Let $(\Omega,\mathcal{H}_T,\mathbb{P})$ be a complete probability space, where $\mathbb{P}$ is a given physical or real--world probability measure. Other elements of the probability space will be defined below. To introduce a regime--switching, we assume that $\{s_t\}_{t=1}^T$ is a homogeneous Markov chain with $N$ state and $\mathsf{P}:=\{p_{ij}\}_{i=0,j=1}^N$ is a random transition probability matrix, including an initial probability vector, where $\{p_{0j}\}_{j=1}^N$ is the initial probability vector. We consider a Bayesian Markov--Switching Vector Autoregressive process of $p$ order (MS--VAR($p$)), which is given by the following equation 
\begin{equation}\label{08001}
y_t=A_{0,s_t}\psi_t+A_{1,s_t}y_{t-1}+\dots +A_{p,s_t}y_{t-p}+\xi_t,~t=1,\dots,T,
\end{equation}
where $y_t=(y_{1,t},\dots,y_{n,t})'$ is an $(n\times 1)$ vector of endogenous variables, $\psi_t=(1,\psi_{2,t},\dots,\psi_{l,t})'$ is an $(l\times 1)$ vector of exogenous variables, $\xi_t=(\xi_{1,t},\dots,\xi_{n,t})'$ is an $(n\times 1)$ residual process, $A_{0,s_t}$ is an $(n\times l)$ random coefficient matrix at regime $s_t$, corresponding to the vector of exogenous variables, for $i=1,\dots,p$, $A_{i,s_t}$ are $(n\times n)$ random coefficient matrices at regime $s_t$, corresponding to $y_{t-1},\dots,y_{t-p}$. Equation \eqref{08001} can be written by
\begin{equation}\label{08002}
y_t=\Pi_{s_t}\mathsf{Y}_{t}+\xi_t,~t=1,\dots,T,
\end{equation}
where $\Pi_{s_t}:=[A_{0,s_t}: A_{1,s_t}:\dots:A_{p,s_t}]$ is an $(n\times d)$ random coefficient matrix with $d:=l+np$ at regime $s_t$, which consist of all the random coefficient matrices and $\mathsf{Y}_{t}:=(\psi_t',y_{t-1}',\dots,y_{t-p}')'$ is a $(d\times 1)$ vector, which consist of exogenous variable $\psi_t$ and last $p$ lagged values of the process $y_t$. The process $\mathsf{Y}_t$ is measurable with respect to a $\sigma$--field $\mathcal{F}_{t-1}$, which is defined below.

For the residual process $\xi_t$, we assume that it has $\xi_t:=\Sigma_{s_t}^{1/2}\varepsilon_t$, $t=1,\dots,T$ representation, see \citeA{Lutkepohl05} and \citeA{McNeil05}, where $\Sigma_{s_t}^{1/2}$ is a Cholesky factor of a positive definite $(n\times n)$ random matrix $\Sigma_{s_t}$, which is measurable with respect to $\sigma$--field $\mathcal{H}_{t-1}$, defined below and depends on $(n_*\times d_*)$ random coefficient matrix $\Gamma_{s_t}:=[B_{0,s_t}:B_{1,s_t}:\dots:B_{p_*+q_*,s_t}]$ with $d_*:=l_*+n_*(p_*+q_*)$. Here $B_{0,s_t}$ is an $(n_*\times l_*)$ random matrix, for $i=1,\dots,p_*+q_*$, $B_{i,s_t}$ are $(n_*\times n_*)$ random matrices, and $\varepsilon_1,\dots,\varepsilon_T$ is a random sequence of independent identically multivariate normally distributed random vectors with means of 0 and covariance matrices of $n$ dimensional identity matrix $I_n$. Then, in particular, for multivariate GARCH process of $(p_*,q_*)$ order, dependence of $\Sigma_{s_t}^{1/2}$ on $\Gamma_{s_t}$ is given by 
\begin{equation}\label{08003}
\text{vech}\big(\Sigma_{s_t}\big)=B_{0,s_t}+\sum_{i=1}^{p_*}B_{i,s_t}\text{vech}\big(\xi_{t-i}\xi_{t-i}'\big)+\sum_{j=1}^{q_*}B_{p_*+j,s_t}\text{vech}(\Sigma_{s_{t-j}}),
\end{equation}
where $B_{0,s_t}$ and $B_{i,s_t}$ for $i=1,\dots, p_*+q_*$ are suitable $([n(n+1)/2]\times 1)$ random vector and suitable $([n(n+1)/2]\times [n(n+1)/2])$ matrices, respectively, and the vech is an operator that stacks elements on and below a main diagonal of a square matrix. Here we assume that initial values of the random covariance matrix are $\Sigma_{s_{1-j}}=\Sigma_{1-j}$ for $j=1,\dots,q_*$.

Let us introduce stacked vectors and matrices: $y:=(y_1',\dots,y_T')'$, $s:=(s_1,\dots,s_T)'$, $\Pi_s:=[\Pi_{s_1}:\dots:\Pi_{s_T}]$, and $\Gamma_s:=[\Gamma_{s_1}:\dots:\Gamma_{s_T}]$. We also assume that the strong white noise process $\{\varepsilon_t\}_{t=1}^T$ is independent of the random coefficient matrices $\Pi_s$ and $\Gamma_s$, random transition matrix $\mathsf{P}$, and regime vector $s$ conditional on initial information $\mathcal{F}_0:=\sigma(y_{1-p}',\dots,y_0',\psi_{1},\dots,\psi_T,\Sigma_{1-q_*},\dots,\Sigma_0)$. Here for a generic random vector $X$, $\sigma(X)$ denotes a $\sigma$--field generated by the random vector $X$, $\Sigma_{1-q_*},\dots,\Sigma_0$ are the initial values of the random matrix process $\Sigma_{s_t}$, $\psi_1,\dots,\psi_T$ are the values of exogenous variables and they are known at time zero. We further suppose that the transition probability matrix $\mathsf{P}$ is independent of the random coefficient matrices $\Pi_s$ and $\Gamma_s$ given initial information $\mathcal{F}_0$ and regime vector $s$.  

To ease of notations, for a generic vector $o=(o_1',\dots,o_T')'$, we denote its first $t$ and last $T-t$ sub vectors by $\bar{o}_t$ and $\bar{o}_t^c$, respectively, that is, $\bar{o}_t:=(o_1',\dots,o_t')'$ and $\bar{o}_t^c:=(o_{t+1}',\dots,o_T')'$. We define $\sigma$--fields: for $t=0,\dots,T$, $\mathcal{F}_{t}:=\mathcal{F}_0\vee\sigma(\bar{y}_{t})$ and $\mathcal{H}_t:=\mathcal{F}_t\vee \sigma(\Pi_s)\vee \sigma(\Gamma_s)\vee \sigma(s)\vee \sigma(\mathsf{P})$ where for generic sigma fields $\mathcal{O}_1$ and $\mathcal{O}_2$, $\mathcal{O}_1\vee \mathcal{O}_2$ is the minimal $\sigma$--field containing the $\sigma$--fields $\mathcal{O}_1$ and $\mathcal{O}_2$. For the first--order Markov chain, a conditional probability that the regime at time $t+1$, $s_{t+1}$ equals some particular value conditional on the past regimes $\bar{s}_t$, transition probability matrix $\mathsf{P}$, and initial information $\mathcal{F}_0$ depends only through the most recent regime at time $t$, $s_t$, transition probability matrix $\mathsf{P}$, and initial information $\mathcal{F}_0$, that is,
\begin{equation}\label{08004}
p_{s_ts_{t+1}}:=\mathbb{P}[s_{t+1}=s_{t+1}|s_t=s_t,\mathsf{P},\mathcal{F}_0]=\mathbb{P}\big[s_{t+1}=s_{t+1}|\bar{s}_t=\bar{s}_t,\mathsf{P},\mathcal{F}_0\big]
\end{equation} 
for $t=0,\dots,T-1$, where $p_{s_1}:=p_{0s_1}=\mathbb{P}[s_1=s_1|\mathsf{P},\mathcal{F}_0]$ is the initial probability. 
A distribution of a residual random vector $\xi:=(\xi_1',\dots,\xi_T')'$ is given by
\begin{equation}\label{08005}
\xi=(\xi_1',\dots,\xi_T')'~|~\mathcal{H}_0\sim \mathcal{N}(0,\Sigma_s),
\end{equation}
where $\Sigma_s:=\text{diag}\{\Sigma_{s_1},\dots,\Sigma_{s_T}\}$ is an $([nT]\times [nT])$ block diagonal matrix. 

To remove duplicates in the random coefficient matrix $(\Pi_s,\Gamma_s)$, for a generic regime vector with length $k$, $o=(o_1,\dots,o_k)'$, we define sets
\begin{equation}\label{08006}
\mathcal{A}_{\bar{o}_t}:=\mathcal{A}_{\bar{o}_{t-1}}\cup\big\{o_t\in \{o_1,\dots,o_k\}\big|o_t\not \in \mathcal{A}_{\bar{o}_{t-1}}\big\},~~~t=1,\dots,k,
\end{equation}
where for $t=1,\dots,k$, $o_t\in \{1,\dots,N\}$ and an initial set is empty set, i.e., $\mathcal{A}_{\bar{o}_0}=\O$. The final set $\mathcal{A}_o=\mathcal{A}_{\bar{o}_k}$ consists of different regimes in regime vector $o=\bar{o}_k$ and $|\mathcal{A}_o|$ represents a number of different regimes in the regime vector $o$. 

Let us assume that elements of sets $\mathcal{A}_s$, $\mathcal{A}_{\bar{s}_t}$, $\mathcal{A}_{\bar{s}_t^c}$, intersection set of the sets $\mathcal{A}_{\bar{s}_t}$ and $\mathcal{A}_{\bar{s}_t^c}$, and difference sets between the sets $\mathcal{A}_{\bar{s}_t^c}$ and $\mathcal{A}_{\bar{s}_t}$ are given by $\mathcal{A}_s=\{\hat{s}_1,\dots,\hat{s}_{r_{\hat{s}}}\}$, $\mathcal{A}_{\bar{s}_t}=\{\alpha_1,\dots,\alpha_{r_\alpha}\}$, $\mathcal{A}_{\bar{s}_t^c}=\{\beta_1,\dots,\beta_{r_\beta}\}$, $\mathcal{A}_{\bar{s}_t}\cap \mathcal{A}_{\bar{s}_t^c}=\{\gamma_1,\dots,\gamma_{r_\gamma}\}$, $\mathcal{A}_{\bar{s}_t^c}\backslash \mathcal{A}_{\bar{s}_t}=\{\delta_1,\dots,\delta_{r_\delta}\}$, and $\mathcal{A}_{\bar{s}_t}\backslash \mathcal{A}_{\bar{s}_t^c}=\{\epsilon_1,\dots,\epsilon_{r_\epsilon}\}$, respectively, where $r_{\hat{s}}:=|\mathcal{A}_s|$, $r_\alpha:=|\mathcal{A}_{\bar{s}_t}|$, $r_\beta:=|\mathcal{A}_{\bar{s}_t^c}|$, $r_\gamma:=|\mathcal{A}_{\bar{s}_t}\cap \mathcal{A}_{\bar{s}_t^c}|$, $r_\delta:=|\mathcal{A}_{\bar{s}_t^c}\backslash \mathcal{A}_{\bar{s}_t}|$, and $r_\epsilon:=|\mathcal{A}_{\bar{s}_t}\backslash \mathcal{A}_{\bar{s}_t^c}|$ are numbers of elements of the sets, respectively. Note that 
\begin{equation}\label{08007}
\mathcal{A}_{\bar{s}_t}=(\mathcal{A}_{\bar{s}_t}\backslash \mathcal{A}_{\bar{s}_t^c})\cup(\mathcal{A}_{\bar{s}_t}\cap \mathcal{A}_{\bar{s}_t^c}),
\end{equation}
\begin{equation}\label{08008}
\mathcal{A}_{\bar{s}_t^c}=(\mathcal{A}_{\bar{s}_t}\cap \mathcal{A}_{\bar{s}_t^c})\cup (\mathcal{A}_{\bar{s}_t^c}\backslash \mathcal{A}_{\bar{s}_t}),
\end{equation}
and
\begin{equation}\label{08009}
\mathcal{A}_s=\mathcal{A}_{\bar{s}_t}\cup \mathcal{A}_{\bar{s}_t^c}=(\mathcal{A}_{\bar{s}_t}\backslash \mathcal{A}_{\bar{s}_t^c})\cup\mathcal{A}_{\bar{s}_t^c}=\mathcal{A}_{\bar{s}_t}\cup(\mathcal{A}_{\bar{s}_t^c}\backslash \mathcal{A}_{\bar{s}_t})
\end{equation}
and intersection sets of the sets of right hand sides of equations \eqref{08007} and \eqref{08008}, and \eqref{08009} are empty sets. We introduce the following regime vectors: $\hat{s}:=(\hat{s}_1,\dots,\hat{s}_{r_{\hat{s}}})'$ is an $(r_{\hat{s}}\times 1)$ vector, $\alpha:=(\alpha_1,\dots,\alpha_{r_\alpha})'$ is an $(r_\alpha\times 1)$ vector, $\beta=(\beta_1,\dots,\beta_{r_\beta})'$ is an $(r_\beta\times 1)$ vector, $\gamma=(\gamma_1,\dots,\gamma_{r_\gamma})'$ is an $(r_\gamma\times 1)$ vector, $\delta=(\delta_1,\dots,\delta_{r_\delta})'$ is an $(r_\delta\times 1)$ vector, and $\epsilon=(\epsilon_1,\dots,\epsilon_{r_\epsilon})'$ is an $(r_\epsilon\times 1)$ vector. For the regime vector $a=(a_1,\dots,a_{r_a})' \in\{\hat{s},\alpha,\beta,\gamma,\delta,\epsilon\}$, we also introduce duplication removed random coefficient matrices, whose block matrices are different:  $\Pi_a=[\Pi_{a_1}:\dots:\Pi_{a_{r_a}}]$ is an $(n\times [dr_a])$ matrix, $\Gamma_a=[\Gamma_{a_1}:\dots:\Gamma_{a_{r_a}}]$ is an $(n_*\times [d_*r_a])$ matrix, and $(\Pi_a,\Gamma_a)$.

We assume that for given duplication removed regime vector $\hat{s}$ and initial information $\mathcal{F}_0$, the coefficient matrices $(\Pi_{\hat{s}_1},\Gamma_{\hat{s}_1}),\dots,(\Pi_{\hat{s}_{r_{\hat{s}}}},\Gamma_{\hat{s}_{r_{\hat{s}}}})$ are independent. Under the last assumption, conditional on $\hat{s}$ and $\mathcal{F}_0$, a joint density function of the random coefficient matrix $(\Pi_{\hat{s}},\Gamma_{\hat{s}})$ is represented by
\begin{equation}\label{08010}
f\big(\Pi_{\hat{s}},\Gamma_{\hat{s}}\big|\hat{s},\mathcal{F}_0\big)=\prod_{t=1}^{r_{\hat{s}}}f\big(\Pi_{\hat{s}_t},\Gamma_{\hat{s}_t}\big|\hat{s}_t,\mathcal{F}_0\big),
\end{equation}
where for a generic random vector $X$, we denote its density function by $f(X)$. Throughout the paper we fix $t=1,\dots,T-1$. Using the regime vectors $\alpha$ and $\delta$, the above joint density function can be written by 
\begin{equation}\label{08011}
f\big(\Pi_{\hat{s}},\Gamma_{\hat{s}}\big|\hat{s},\mathcal{F}_0\big)=
f\big(\Pi_{\alpha},\Gamma_{\alpha}\big|\alpha,\mathcal{F}_0\big)f_*\big(\Pi_{\delta},\Gamma_{\delta}\big|\delta,\mathcal{F}_0\big)
\end{equation}
where the density function $f_*\big(\Pi_{\delta},\Gamma_{\delta}\big|\delta,\mathcal{F}_0\big)$ equals
\begin{equation}\label{08012}
f_*\big(\Pi_\delta,\Gamma_\delta\big|\delta,\mathcal{F}_0\big):=
\begin{cases}
f\big(\Pi_\delta,\Gamma_\delta\big|\delta,\mathcal{F}_0\big),& \text{if}~~~r_\delta\neq 0,\\
1,& \text{if}~~~r_\delta= 0.
\end{cases}
\end{equation}
Then, the following Proposition, which is useful for Monte--Carlo simulation holds, see below. The proofs of this one and other Propositions are given in Appendix.
 
\begin{proposition}\label{prop01}
Conditional on initial information $\mathcal{F}_0$, a joint density function of the random vectors $\bar{y}_t$ and $s$ and random matrices $\Pi_{\hat{s}}$, $\Gamma_{\hat{s}}$, and $\mathsf{P}$ is given by
\begin{eqnarray}\label{08013}
f(\bar{y}_t,\Pi_{\hat{s}},\Gamma_{\hat{s}},s,\mathsf{P}|\mathcal{F}_0)=f\big(\bar{y}_t,\Pi_{\alpha},\Gamma_{\alpha},\bar{s}_t\big|\mathcal{F}_0\big)f_*\big(\Pi_{\delta},\Gamma_{\delta}\big|\delta,\mathcal{F}_0\big)f(s,\mathsf{P}|\mathcal{F}_0)\big/f(\bar{s}_t|\mathcal{F}_0).
\end{eqnarray}
In particular, the following relationships holds
\begin{eqnarray}\label{08018}
f(\bar{s}_t^c|\Pi_\alpha,\Gamma_\alpha,\bar{s}_t,\mathsf{P},\mathcal{F}_t)=f(\bar{s}_t^c|\bar{s}_t,\mathsf{P},\mathcal{F}_0),
\end{eqnarray}
\begin{equation}\label{08015}
f\big(\Pi_{\delta},\Gamma_{\delta}\big|\Pi_{\alpha},\Gamma_{\alpha},s,\mathsf{P},\mathcal{F}_t\big)=f_*\big(\Pi_{\delta},\Gamma_{\delta}\big|\delta,\mathcal{F}_0\big),
\end{equation}
\begin{equation}\label{08016}
f\big(\Pi_{\beta},\Gamma_{\beta}\big|s,\mathcal{F}_t\big)=f\big(\Pi_{\gamma},\Gamma_{\gamma}\big|\bar{s}_t,\mathcal{F}_t\big)f_*\big(\Pi_{\delta},\Gamma_{\delta}\big|\delta,\mathcal{F}_0\big),
\end{equation}
\begin{eqnarray}\label{08017}
f(\mathsf{P}|\bar{s}_t,\Pi_\alpha,\Gamma_\alpha,\mathcal{F}_t)=f(\mathsf{P}|\bar{s}_t,\mathcal{F}_0),
\end{eqnarray}
and
\begin{equation}\label{08019}
f(\Pi_\alpha,\Gamma_\alpha|\bar{s}_t,\mathsf{P},\mathcal{F}_t)=f(\Pi_\alpha,\Gamma_\alpha|\bar{s}_t,\mathcal{F}_t).
\end{equation}
\end{proposition}

It follows from the Proposition that (i) conditional on $\bar{s}_t$, $\mathsf{P}$, and $\mathcal{F}_0$, $\bar{s}_t^c$ and $(\bar{y}_t,\Pi_\alpha,\Gamma_\alpha)$ are independent, (ii) conditional on $\delta$ and $\mathcal{F}_0$, $(\Pi_{\delta},\Gamma_{\delta})$ and $(\bar{y}_t,\Pi_{\alpha},\Gamma_{\alpha},\alpha,\mathsf{P})$ are independent, (iii) conditional on $\bar{s}_t$ and $\mathcal{F}_0$, $\mathsf{P}$ and $(\bar{y}_t,\Pi_\alpha,\Gamma_\alpha)$ are independent, and (iv) conditional on $\bar{s}_t$ and $\mathcal{F}_t$, $(\Pi_\alpha,\Gamma_\alpha)$ and $\mathsf{P}$ are independent. It should be noted that according to the Markov property \eqref{08004}, it follows from equation \eqref{08018} that the assumption for a Markov chain in the book of \citeA{Hamilton94} always holds, namely,
\begin{eqnarray}\label{08156}
f(s_{t+1}|\Pi_\alpha,\Gamma_\alpha,\bar{s}_t,\mathsf{P},\mathcal{F}_t)=f(s_{t+1}|s_t,\mathsf{P},\mathcal{F}_0).
\end{eqnarray}

\section{Special Case of Bayesian MS--VAR process}

In this section, we consider a special case of the Bayesian MS--VAR($p$) process. The Bayesian MS--VAR($p$) process can be written by the following equation  
\begin{equation}\label{08030}
y_t=\Pi_{s_t}\mathsf{Y}_{t}+\xi_t=(\mathsf{Y}_{t}'\otimes I_n)\pi_{s_t}+\xi_t, ~~~t=1,\dots,T,
\end{equation}
where $\otimes$ is the Kronecker product of two matrices and $\pi_{s_t}:=\text{vec}(\Pi_{s_t})$ is an $(nd\times 1)$ vectorization of the random coefficient matrix $\Pi_{s_t}$. Now we define distributions of the random coefficient vector $\pi_{s_t}$ and covariance matrix $\Sigma_{s_t}$. We assume that conditional on the regime $s_t$ and initial information $\mathcal{F}_0$, a distribution of the random covariance matrix $\Sigma_{s_t}$ is given by 
\begin{equation}\label{08031}
\Sigma_{s_t}~|~s_t,\mathcal{F}_0\sim \mathcal{IW}(\nu_{0,s_t},V_{0,s_t})
\end{equation}
where the notation $\mathcal{IW}$ denotes the Inverse--Wishart distribution, $\nu_{0,s_t}>n-1$ is a degrees of freedom and $V_{0,s_t}$ is a positive definite scale matrix and both are prior hyperparameters, corresponding to the regime $s_t$. Consequently, a distribution of the residual vector $\xi_t$ equals 
\begin{equation}\label{08032}
\xi_t~|~\Sigma_{s_t},s_t,\mathcal{F}_0\sim \mathcal{N}\Big(0,\Sigma_{s_t}\Big),
\end{equation}
where $\mathcal{N}$ denotes the normal distribution. Also, we assume that conditional on the covariance matrix $\Sigma_{s_t}$, regime $s_t$, and initial information $\mathcal{F}_0$, a distribution of the random coefficient vector $\pi_{s_t}$ is given by 
\begin{equation}\label{08033}
\pi_{s_t}~|~\Sigma_{s_t},s_t,\mathcal{F}_0\sim \mathcal{N}\Big(\pi_{0,s_t},\Lambda_{0,s_t}\otimes \Sigma_{s_t}\Big),
\end{equation}
where $\pi_{0,s_t}$ is an $(nd\times 1)$ prior hyperparameter vector at regime $s_t$ and $\Lambda_{0,s_t}$ is a symmetric positive definite $(d\times d)$ prior hyperparameter matrix at regime $s_t$. 

\subsection{Distributions}

For the regime vector $\bar{s}_t$ and regime $\alpha_k$, we define sets
\begin{equation}\label{08034}
S_{t,\alpha_k}:=\big\{u\in\{1,\dots,t\}\big|s_u=\alpha_k,~u=1,\dots,t\big\},~~~k=1,\dots,r_{\alpha}.
\end{equation}
For $k=1,\dots,r_{\alpha}$, the set $S_{t,\alpha_k}$ consists of indexes of regimes in the regime vector $\bar{s}_t$ that equal the regime $\alpha_k$. Let us suppose that $q_{t,\alpha_k}:=|S_{t,\alpha_k}|$ is a number of regimes in the regime vector $\bar{s}_t$ that equal the regime $\alpha_k$ and elements of the set $S_{t,\alpha_k}$ are given by
\begin{equation}\label{08035}
S_{t,\alpha_k}=\big\{k_{t,1},\dots,k_{t,q_{t,\alpha_k}}\big\},~~~k=1,\dots,r_{\alpha}.
\end{equation}
Further, we define indexes
\begin{equation}\label{08036}
o_t:=\big\{k\in\{1,\dots,r_\alpha\}\big|s_t=\alpha_k,~k=1,\dots,r_\alpha\big\},~~~t=1,\dots,t.
\end{equation}
The index $o_t$ represents a position of the regime $s_t$ in the regime vector $\alpha$. Let $\pi_{\alpha}:=\big(\pi_{\alpha_1}',\dots,\pi_{\alpha_{r_{\alpha}}}'\big)'$ be an $([ndr_{\alpha}]\times 1)$ duplication removed random coefficient vector, whose sub--vectors are different and which corresponds to the regime vector $\bar{s}_t$, $y_{t,\alpha_k}:=\Big(y_{k_{t,1}}',\dots,y_{k_{t,q_{t,\alpha_k}}}'\Big)'$ be an $([nq_{t,\alpha_k}]\times 1)$ vector of endogenous variables, corresponding to the regime $\alpha_k$, and $\mathsf{Y}_{t,\alpha_k}^\circ:=\big[\mathsf{Y}_{k_{t,1}}:\dots:\mathsf{Y}_{k_{t,q_{t,\alpha_k}}}\big]$ be a $(d\times q_{t,\alpha_k})$ matrix of exogenous and endogenous variables, corresponding to the regime $\alpha_k$. By using a $(t\times r_\alpha)$ matrix $D_{\alpha}:=[j_{o_1}:\dots:j_{o_t}]'$, one can revive the vector $\pi_{\bar{s}_t}:=\text{vec}(\Pi_{\bar{s}_t})$ from the vector $\pi_{\alpha}$, that is, $\pi_{\bar{s}_t}=(D_{\alpha}\otimes I_{nd})\pi_{\alpha}$, where $j_o$ is an $(r_\alpha\times 1)$ unit vector, whose $o$--th element equals one and others zero. 

\subsubsection{Conditional Densities}

It follows from equations \eqref{08032} and \eqref{08033} that distributions of the random vectors $\bar{\xi}_t$ and $\pi_{\alpha}$ are obtained by
\begin{equation}\label{08037}
\bar{\xi}_t~|~\Sigma_{\alpha},\bar{s}_t,\mathcal{F}_0\sim \mathcal{N}\big(0,\Sigma_{\bar{s}_t}\big),
\end{equation}
and
\begin{equation}\label{08038}
\pi_{\alpha}~|~\Sigma_{\alpha},\alpha,\mathcal{F}_0\sim \mathcal{N}\Big(\pi_{0,\alpha},\Sigma_{\pi_{\alpha}}\Big),
\end{equation}
respectively, where $\Sigma_{\bar{s}_t}:=\text{diag}\{\Sigma_{s_1},\dots,\Sigma_{s_t}\}$ is an ($[nt]\times [nt]$) block diagonal matrix, corresponding to the regime vector $\bar{s}_t$ and $\Sigma_{\alpha}:=\big[\Sigma_{\alpha_1}:\dots:\Sigma_{\alpha_{r_{\alpha}}}\big]'$ is an $([nr_{\alpha}]\times n)$ matrix, $\pi_{0,\alpha}:=\big(\pi_{0,\alpha_1}',\dots,\pi_{0,\alpha_{r_{\alpha}}}'\big)'$ is an $([ndr_{\alpha}]\times 1)$ prior hyperparameter vector, and $\Sigma_{\pi_{\alpha}}:=\text{diag}\big\{\Lambda_{0,\alpha_1}\otimes \Sigma_{\alpha_1},\dots,\Lambda_{0,\alpha_{r_{\alpha}}}\otimes \Sigma_{\alpha_{r_{\alpha}}}\big\}$ is an $([ndr_{\alpha}]\times [ndr_{\alpha}])$ block diagonal matrix, all of which correspond to the duplication removed regime vector $\alpha$. A connection between the random matrices $\Sigma_{\bar{s}_t}$ and $\Sigma_{\alpha}$ is
\begin{equation}\label{08039}
\Sigma_{\bar{s}_t}=\text{diag}\big\{\big((D_{\alpha}\otimes I_n)\Sigma_{\alpha}\big)_1,\dots,\big((D_{\alpha}\otimes I_n)\Sigma_{\alpha}\big)_t\big\},
\end{equation}
where the matrix $\big((D_{\alpha}\otimes I_n)\Sigma_{\alpha}\big)_j$ equals $j$--th block matrix of the matrix $(D_{\alpha}\otimes I_n)\Sigma_{\alpha}$. On the other hand, by following \citeA{Battulga24a}, a distribution of the $([nT]\times 1)$ random vector $y=(y_1',\dots,y_T')'$ is given by
\begin{equation}\label{08040}
y~\big|~\pi_{\hat{s}},\Sigma_{\hat{s}},s,\mathcal{F}_0\sim \mathcal{N}\Big(\Psi_s^{-1}\varphi_s,\Psi_s^{-1}\Sigma_s(\Psi_s^{-1})'\Big),
\end{equation}
where the matrix $\Psi_s$ and the vector $\varphi_s$ are
\begin{equation}\label{08041}
\Psi_s:=\begin{bmatrix}
I_n & 0 & \dots & 0 & \dots & 0 & 0\\
-A_{1,s_2} & I_n & \dots & 0 & \dots & 0 & 0\\
\vdots & \vdots & \dots & \vdots & \dots & \vdots & \vdots\\
0 & 0 & \dots & -A_{p-1,s_{T-1}} & \dots & I_n & 0\\
0 & 0 & \dots & -A_{p,s_T} & \dots & -A_{1,s_T} & I_n
\end{bmatrix}
\end{equation}
and
\begin{equation}\label{08042}
\varphi_s:=\begin{bmatrix}
A_{0,s_1}\psi_1+A_{1,s_1}y_{0}+\dots+A_{p,s_1}y_{1-p}\\ 
A_{0,s_2}\psi_2+A_{2,s_2}y_{0}+\dots+A_{p,s_2}y_{2-p}\\
\vdots\\
A_{0,s_{T-1}}\psi_{T-1}\\
A_{0,s_T}\psi_T
\end{bmatrix},
\end{equation}
respectively. To price default--free options, \citeA{Battulga24a} used the conditional distribution of the random vector $y$. For a generic vector $o=(o_1',\dots,o_n')'$ with ($m\times 1$) vector $o_i$, we introduce an $(m\times n)$ matrix notation $o^\circ:=[o_1:\dots:o_n]$. Then, the following Proposition holds. 

\begin{proposition}\label{prop02}
Let for $t=1,\dots,T-1$, $\pi_{s_t}~|~\Sigma_{s_t},s_t,\mathcal{F}_0\sim \mathcal{N}\big(\pi_{0,s_t},\Lambda_{0,s_t}\otimes \Sigma_{s_t}\big)$, and $\Sigma_{s_t}~|~s_t,\mathcal{F}_0\sim \mathcal{IW}(\nu_{0,s_t},V_{0,s_t})$. Then, first, conditional on the regime vector $\bar{s}_t$ and initial information $\mathcal{F}_0$, a joint density function of the random vector $\bar{y}_t$ is given by
\begin{eqnarray}\label{08043}
f(\bar{y}_t|\bar{s}_t,\mathcal{F}_0)=\frac{1}{\pi^{nt/2}}\prod_{k=1}^{r_{\alpha}}\frac{|\Lambda_{0,\alpha_k}^{-1}|^{n/2}\Gamma_{n}\big(\nu_{0,\alpha_k|t}/2\big)|V_{0,\alpha_k}|^{\nu_{0,\alpha_k}/2}}{|\Lambda_{0,\alpha_k|t}^{-1}|^{n/2}\Gamma_{n}(\nu_{0,\alpha_k}/2)\big|B_{t,\alpha_k}+V_{0,\alpha_k}\big|^{\nu_{0,\alpha_k|t}/2}},
\end{eqnarray}
where $\Gamma_n(\cdot)$ is the multivariate gamma function, $\Lambda_{0,\alpha_k|t}^{-1}:=\mathsf{Y}_{t,\alpha_k}^\circ(\mathsf{Y}_{t,\alpha_k}^\circ)'+\Lambda_{0,\alpha_k}^{-1}$ is a $(d\times d)$ matrix, and $B_{t,\alpha_k}$ is an  $(n\times n)$ positive semi--definite matrix and equals 
\begin{eqnarray}\label{08044}
B_{t,\alpha_k}&:=&y_{t,\alpha_k}^{\circ}(y_{t,\alpha_k}^{\circ})'+\pi_{0,\alpha_k}^{\circ}\Lambda_{0,\alpha_k}^{-1}(\pi_{0,\alpha_k}^{\circ})'-\pi_{0,\alpha_k|t}^\circ\Lambda_{0,\alpha_k|t}^{-1}(\pi_{0,\alpha_k|t}^\circ)'\\
&=&\big(y_{t,\alpha_k}^{\circ}-\pi_{0,\alpha_k}^{\circ}\mathsf{Y}_{t,\alpha_k}^\circ\big)\big(I_{q_{t,\alpha_k}}+(\mathsf{Y}_{t,\alpha_k}^\circ)'\Lambda_{0,\alpha_k}\mathsf{Y}_{t,\alpha_k}^\circ\big)^{-1}\big(y_{t,\alpha_k}^{\circ}-\pi_{0,\alpha_k}^{\circ}\mathsf{Y}_{t,\alpha_k}^\circ\big)'\nonumber
\end{eqnarray}
with $(n\times d)$ matrix $\pi_{0,\alpha_k|t}^\circ:=\big(y_{t,\alpha_k}^{\circ}(\mathsf{Y}_{t,\alpha_k}^\circ)'+\pi_{0,\alpha_k}^{\circ}\Lambda_{0,\alpha_k}^{-1}\big)\Lambda_{0,\alpha_k|t}$. Second, conditional on the random covariance matrix $\Sigma_{\alpha}$, regime vector $\bar{s}_t$, and information $\mathcal{F}_t$, a joint density function of the random coefficient vector $\pi_\alpha$ is given by
\begin{eqnarray}\label{08045}
&&f(\pi_{\alpha}|\Sigma_{\alpha},\bar{s}_t,\mathcal{F}_t)\\
&&= \frac{1}{(2\pi)^{ndr_{\alpha}/2}\prod_{k=1}^{r_{\alpha}}|A_{\alpha_k|t}|^{1/2}} \exp\bigg\{-\frac{1}{2}\sum_{k=1}^{r_{\alpha}}\Big(\pi_{\alpha_k}-\pi_{0,\alpha_k|t}\Big)'A_{\alpha_k|t}^{-1}\Big(\pi_{\alpha_k}-\pi_{0,\alpha_k|t}\Big)\bigg\},\nonumber
\end{eqnarray}
where for $k=1,\dots,r_{\alpha}$, $A_{\alpha_k|t}:=\big(\Lambda_{0,\alpha_k|t}\otimes \Sigma_{\alpha_k}\big)$ is an $([nd]\times [nd])$ matrix and $\pi_{0,\alpha_k|t}:=\big((\Lambda_{0,\alpha_k|t}\mathsf{Y}_{t,\alpha_k}^\circ)\otimes I_n\big)y_{t,\alpha_k}+\big((\Lambda_{0,\alpha_k|t}\Lambda_{0,\alpha_k}^{-1})\otimes I_n\big)\pi_{0,\alpha_k}$ is an $([nd]\times 1)$ vector. Third, conditional on the regime vector $\bar{s}_t$ and information $\mathcal{F}_t$, a joint density function of the random coefficient matrix $\Sigma_{\alpha}$ is given by
\begin{eqnarray}\label{08046}
f(\Sigma_{\alpha}|\bar{s}_t,\mathcal{F}_t)&=&\prod_{k=1}^{r_{\alpha}}\frac{\big|B_{t,\alpha_k}+V_{0,\alpha_k}\big|^{\nu_{0,\alpha_k|t}/2}}{\Gamma_{n}\big(\nu_{0,\alpha_k|t}/2\big)2^{n\nu_{0,\alpha_k|t}/2}}|\Sigma_{\alpha_k}|^{-(\nu_{0,\alpha_k|t}+n+1)/2}\nonumber\\
&\times&\exp\bigg\{-\frac{1}{2}\sum_{k=1}^{r_{\alpha}}\mathrm{tr}\Big(\big(B_{t,\alpha_k}+V_{0,\alpha_k}\big)\Sigma_{\alpha_k}^{-1}\Big)\bigg\},
\end{eqnarray}
where $\nu_{0,\alpha_k|t}:=\nu_{0,\alpha_k}+q_{t,\alpha_k}$. Fourth, conditional on the regime vector $s$ and information $\mathcal{F}_t$, a joint density function of the random coefficient matrix $\pi_{\beta}^\circ$ is given by
\begin{eqnarray}\label{08047}
f(\pi_{\beta}^\circ|s,\mathcal{F}_t)&=&\prod_{k=1}^{r_{\gamma}}\frac{|\Lambda_{0,\gamma_k|t}|^{-n/2}\big|B_{t,\gamma_k}+V_{0,\gamma_k}\big|^{-d/2}\Gamma_{n}\big((\nu_{0,\gamma_k|t}+d)/2\big)}{\pi^{nd/2}\Gamma_{n}(\nu_{0,\gamma_k|t}/2)}\nonumber\\
&\times&\big|I_n+(B_{t,\gamma_k}+V_{0,\gamma_k})^{-1}(\pi_{\gamma_k}^\circ-\pi_{0,\gamma_k|t}^\circ)\Lambda_{0,\gamma_k|t}^{-1}(\pi_{\gamma_k}^\circ-\pi_{0,\gamma_k|t}^\circ)'\big|^{-(\nu_{0,\gamma_k|t}+d)/2}\nonumber\\
&\times&\prod_{\ell=1}^{r_{\delta}}\frac{|\Lambda_{0,\delta_\ell}|^{-n/2}|V_{0,\delta_\ell}|^{-d/2}\Gamma_{n}\big((\nu_{0,\delta_\ell}+d)/2\big)}{\pi^{nd/2}\Gamma_{n}(\nu_{0,\delta_\ell}/2)}\\
&\times&\big|I_n+V_{0,\delta_\ell}^{-1}(\pi_{\delta_\ell}^\circ-\pi_{0,\delta_\ell}^\circ)\Lambda_{0,\delta_\ell}^{-1}(\pi_{\delta_\ell}^\circ-\pi_{0,\delta_\ell}^\circ)'\big|^{-(\nu_{0,\delta_\ell}+d)/2}.\nonumber
\end{eqnarray}
Finally, an $(n\times n)$ matrix $B_{T,\alpha_k}$ is represented by
\begin{eqnarray}\label{08162}
B_{T,\alpha_k}&=&B_{t,\alpha_k}+\big(y_{t,\alpha_k}^*-\pi_{0,\alpha_k}^\circ\mathsf{Y}_{t,\alpha_k}^*-(y_{t,\alpha_k}^\circ-\pi_{0,\alpha_k}^\circ\mathsf{Y}_{t,\alpha_k}^\circ)\Phi_{t,\alpha_k}(\mathsf{Y}_{t,\alpha_k}^\circ)'\Lambda_{0,\alpha_k}\mathsf{Y}_{t,\alpha_k}^*\big)\nonumber\\
&\times& (I_{q_{t,\alpha_k}^*}+\mathsf{Y}_{t,\alpha_k}^*\Lambda_{0,\alpha_k|t} \mathsf{Y}_{t,\alpha_k}^*)^{-1}\nonumber\\
&\times&\big(y_{t,\alpha_k}^*-\pi_{0,\alpha_k}^\circ\mathsf{Y}_{t,\alpha_k}^*-(y_{t,\alpha_k}^\circ-\pi_{0,\alpha_k}^\circ\mathsf{Y}_{t,\alpha_k}^\circ)\Phi_{t,\alpha_k}(\mathsf{Y}_{t,\alpha_k}^\circ)'\Lambda_{0,\alpha_k}\mathsf{Y}_{t,\alpha_k}^*\big)',
\end{eqnarray}
where the matrices $y_{t,\alpha_k}^*$ and $\mathsf{Y}_{t,\alpha_k}^*$ come from matrices $y_{T,\alpha_k}^\circ=[y_{t,\alpha_k}^\circ:y_{t,\alpha_k}^*]$ and $\mathsf{Y}_{T,\alpha_k}^\circ=[\mathsf{Y}_{t,\alpha_k}^\circ:\mathsf{Y}_{t,\alpha_k}^*]$ and $q_{t,\alpha_k}^*=q_{T,\alpha_k}-q_{t,\alpha_k}$. If $q_{T,\alpha_k}=q_{t,\alpha_k}$, then $B_{T,\alpha_k}=B_{t,\alpha_k}$.
\end{proposition}

It follows from equations \eqref{08045} and \eqref{08046} that sub coefficient vectors and sub covariance matrices are conditional independent. Note that the conditional independence is consistent with the assumption \eqref{08010}. From equations \eqref{08045} and \eqref{08046} one deduces that for $k=1,\dots,r_{\alpha}$, the conditional density functions of the coefficient vector $\pi_{\alpha_k}$ and the covariance matrix $\Sigma_{\alpha_k}$ are given by
\begin{equation}\label{08067}
f(\pi_{\alpha_k}|\Sigma_{\alpha_k},\alpha_k,\mathcal{F}_t)= \frac{1}{(2\pi)^{nd/2}|A_{\alpha_k|t}|^{1/2}} \exp\bigg\{-\frac{1}{2}\Big(\pi_{\alpha_k}-\pi_{0,\alpha_k|t}\Big)'A_{\alpha_k|t}^{-1}\Big(\pi_{\alpha_k}-\pi_{0,\alpha_k|t}\Big)\bigg\}
\end{equation}
and
\begin{eqnarray}\label{08068}
f(\Sigma_{\alpha_k}|\alpha_k,\mathcal{F}_t)&=&\frac{\big|B_{t,\alpha_k}+V_{0,\alpha_k}\big|^{\nu_{0,\alpha_k|t}/2}}{\Gamma_{n}\big(\nu_{0,\alpha_k|t}/2\big)2^{n\nu_{0,\alpha_k|t}/2}}|\Sigma_{\alpha_k}|^{-(\nu_{0,\alpha_k|t}+n+1)/2}\nonumber\\
&\times&\exp\bigg\{-\frac{1}{2}\mathrm{tr}\Big(\big(B_{t,\alpha_k}+V_{0,\alpha_k}\big)\Sigma_{\alpha_k}^{-1}\Big)\bigg\},
\end{eqnarray}
respectively. Thus, the conditional distribution functions of the coefficient vector $\pi_{\alpha_k}$ and the covariance matrix $\Sigma_{\alpha_k}$ are multivariate normal and inverse Wishart, respectively. Also, it follows from equation \eqref{08047} that the conditional density function of the random coefficient matrix $\pi_\beta^\circ$ equals products of matrix variate student $t$ density functions. The conditional density function of the random matrix $\pi_\beta^\circ$ can be used to impulse response analysis. Because marginal density functions of the random coefficient matrix $\pi_\beta^\circ$ are the matrix variate student $t$, their means are given by
\begin{equation}\label{08069}
\mathbb{E}\big[\pi_{\gamma_k}^\circ\big|\gamma_k,\mathcal{F}_t\big]=\pi_{0,\gamma_k|t}^\circ,~~~k=1,\dots,r_\gamma
\end{equation}
and
\begin{equation}\label{08070}
\mathbb{E}\big[\pi_{\delta_k}^\circ\big|\delta_k,\mathcal{F}_0\big]=\pi_{0,\delta_k}^\circ,~~~k=1,\dots,r_\delta.
\end{equation}

Because, according to equation \eqref{08044}, density function \eqref{08043} has a form of the matrix variate student $t$ distribution, we refer to the density function as a conditional matrix variate student $t$ density function. Furthermore, it follows from equation \eqref{08009} and \eqref{08043} that conditional on the regime vector $s$ and information $\mathcal{F}_t$, a density function of future values of the vector of endogenous variables is given by the following equation
\begin{eqnarray}\label{08071}
f(\bar{y}_t^c|s,\mathcal{F}_t)&=&\frac{1}{\pi^{n(T-t)/2}}\prod_{k=1}^{r_{\alpha}}\frac{|\Lambda_{0,\alpha_k|t}^{-1}|^{n/2}\Gamma_{n}\big(\nu_{0,\alpha_k|T}/2\big)|B_{t,\alpha_k}+V_{0,\alpha_k}|^{\nu_{0,\alpha_k|t}/2}}{|\Lambda_{0,\alpha_k|T}^{-1}|^{n/2}\Gamma_{n}(\nu_{0,\alpha_k|t}/2)\big|B_{T,\alpha_k}+V_{0,\alpha_k}\big|^{\nu_{0,\alpha_k|T}/2}}\nonumber\\
&\times&\prod_{\ell=1}^{r_{\delta}}\frac{|\Lambda_{0,\delta_\ell}^{-1}|^{n/2}\Gamma_{n}\big(\nu_{0,\delta_\ell|T}/2\big)|V_{0,\delta_\ell}|^{\nu_{0,\delta_\ell}/2}}{|\Lambda_{0,\delta_\ell|T}^{-1}|^{n/2}\Gamma_{n}(\nu_{0,\delta_\ell}/2)\big|B_{T,\delta_\ell}+V_{0,\delta_\ell}\big|^{\nu_{0,\delta_\ell|T}/2}}.
\end{eqnarray}
Consequently, due to equation \eqref{08162}, the above density function  is represented by a product of the conditional matrix variate student $t$ density functions. Note that if $r_\delta=0$, we must eliminate the second line of the above equation.

Let us assume that the prior density functions of each row of the transition probability matrix $\mathsf{P}$ follow Dirichlet distribution and they are mutually independent. Under the assumption, a joint density function of them is given by
\begin{equation}\label{08074}
f(\mathsf{P}|\mathcal{F}_0)=\prod_{i=0}^N\frac{\Gamma\big(\sum_{j=1}^N\alpha_{ij}\big)}{\prod_{j=1}^N\Gamma(\alpha_{ij})}\prod_{j=1}^Np_{ij}^{\alpha_{ij}-1}
\end{equation}
where $\Gamma(\cdot)$ is the gamma function and the parameters of Dirichlet distribution satisfy $\alpha_{ij}>0$ for $i=0,\dots,N$ and $j=1,\dots,N$. Let us denote $i$--th row of the random transition probability matrix $\mathsf{P}$ by $\mathsf{P}_i$, corresponding prior hyperparameter by $\alpha_i:=(\alpha_{i1},\dots,\alpha_{iN})'$, and Dirichlet distribution by $\text{Dir}(\alpha_i)$. Then, the following Lemma holds.

\begin{proposition}\label{prop03}
Let for $i=0,\dots,N$, $\mathsf{P}_i\sim\mathrm{Dir}(\alpha_i)$ and they are mutually independent. Then, the followings are hold
\begin{itemize}
\item[(i)] for $t=1,\dots,T$, conditional on the information $\mathcal{F}_0$, a density function of regime vector $\bar{s}_t$ is given by
\begin{equation}\label{08075}
f(\bar{s}_t|\mathcal{F}_0)=\prod_{i=0}^N\frac{\Gamma\big(\sum_{j=1}^N\alpha_{ij}\big)}{\prod_{j=1}^N\Gamma(\alpha_{ij})}\frac{\prod_{j=1}^N\Gamma(\alpha_{ij}+n_{ij}(\bar{s}_t))}{\Gamma\big(\sum_{j=1}^N(\alpha_{ij}+n_{ij}(\bar{s}_t))\big)}
\end{equation}
\item[(ii)] and for $t=2,\dots,T$, conditional on the regime vector $\bar{s}_{t-1}$ and information $\mathcal{F}_0$, a density function of regime $s_t$ is given by
\begin{equation}\label{08076}
f(s_t|\bar{s}_{t-1},\mathcal{F}_0)=\frac{\alpha_{s_{t-1}s_t}+n_{s_{t-1}s_t}(\bar{s}_{t-1})}{\sum_{s_t=1}^N\big(\alpha_{s_{t-1}s_t}+n_{s_{t-1}s_t}(\bar{s}_{t-1})\big)},
\end{equation}
\end{itemize}
where the random variable $n_{ij}(\bar{s}_t)$ equals 
\begin{equation}\label{08077}
n_{ij}(\bar{s}_t):=\#\big\{m\in\{0,1,\dots,t-1\}\big| s_{m-1}=i,s_m=j,~m=2,\dots,t\big\}
\end{equation}
for $t=2,\dots,T$, $i=1,\dots,N$, and $j=1,\dots,N$ and
\begin{equation}\label{08078}
n_{ij}(s_1):=\begin{cases}
1 & \mathrm{if}~~~i=0,~s_1=j\\
0 & \mathrm{if}~~~\mathrm{otherwise}
\end{cases}
\end{equation}
for $i=0,\dots,N$ and $j=1,\dots,N$.
\end{proposition}

Consequently, it is worth mentioning that according to equation \eqref{08076} in the above Proposition, conditional on $\mathcal{F}_0$, the regime--switching process $s_t$ is not a Markov chain because the conditional density function depends on the regime vector $\bar{s}_{t-1}$.

\subsubsection{Characteristic Function}

For $k=1,\dots,r_\gamma$, equation \eqref{08067} can be written by
\begin{eqnarray}\label{08084}
f(\pi_{\gamma_k}^\circ|\Sigma_{\gamma_k},\gamma_k,\mathcal{F}_t)&=& \frac{1}{(2\pi)^{nd/2}|\Lambda_{0,\gamma_k|t}|^{n/2}|\Sigma_{\gamma_k}|^{d/2}}\nonumber\\
&\times& \exp\bigg\{-\frac{1}{2}\mathrm{tr}\Big(\pi_{\gamma_k}^\circ-\pi_{0,\gamma_k|t}^\circ)\Lambda_{0,\gamma_k|t}^{-1}(\pi_{\gamma_k}^\circ-\pi_{0,\gamma_k|t}^\circ)'\Sigma_{\gamma_k}^{-1}\Big)\bigg\}.
\end{eqnarray}
To obtain characteristic function of the random coefficient matrix $\pi_{\gamma_k}^\circ$ for given regime $\gamma_k$ and information $\mathcal{F}_t$, we use the matrix generalized inverse Gaussian (MGIG) distribution. For a positive definite $(n\times n)$ matrix $\Sigma$, the density function of the MGIG distribution is given by
\begin{equation}\label{08085}
f(\Sigma)=\frac{2^{n\lambda}}{|\mathsf{A}|^{\lambda}\mathcal{B}_{\lambda}\big(\frac{1}{4}\mathsf{B}\mathsf{A}\big)}|\Sigma|^{\lambda-(n+1)/2}\exp\bigg\{-\frac{1}{2}\mathrm{tr}\big(\mathsf{A}\Sigma^{-1}+\mathsf{B}\Sigma\big)\bigg\},
\end{equation}
where $\mathcal{B}_\lambda(\cdot)$ is the matrix argument modified Bessel function of the second kind with index $\lambda$, which is defined by
\begin{equation}\label{08155}
\mathcal{B}_\lambda\bigg(\frac{1}{4}\mathsf{B}\mathsf{A}\bigg):=\bigg|\frac{1}{2}\mathsf{B}\bigg|^{-\lambda}\int_{\Sigma>0}|\Sigma|^{-\lambda-(n+1)/2}\exp\bigg\{-\frac{1}{2}\mathrm{tr}\big(\mathsf{A}\Sigma^{-1}+\mathsf{B}\Sigma\big)\bigg\}d\Sigma
\end{equation}
and for $n\geq 2$, the index $\lambda\in \mathbb{R}$ and the $(n\times n)$ matrices $\mathsf{A}$ and $\mathsf{B}$ satisfy
\begin{equation}\label{08086}
\begin{cases}
\mathsf{A}\geq 0,~\mathsf{B}>0 & \text{if}~~~\lambda\geq \frac{1}{2}\\
\mathsf{A}>0,~\mathsf{B}>0 & \text{if}~~~-\frac{1}{2}(n-1)\leq \lambda< \frac{1}{2}\\
\mathsf{A}> 0,~\mathsf{B}\geq 0 & \text{if}~~~\lambda< -\frac{1}{2}(n-1)\\
\end{cases},
\end{equation}
see \citeA{Butler98}. Its one dimensional version is called generalized inverse Gaussian distribution and it is widely used to model returns of financial assets, see \citeA{McNeil05}. It is the well--known fact that a characteristic function of the random coefficient matrix $\pi_{\gamma_k}^\circ$ for given regime $\gamma_k$, covariance matrix $\Sigma_{\gamma_k}$, and information $\mathcal{F}_t$ is given by
\begin{eqnarray}\label{08087}
\varphi(Z_{\gamma_k}|\Sigma_{\gamma_k},\gamma_k,\mathcal{F}_t)&=&\mathbb{E}\big[\exp\big\{\mathrm{tr}\big(\mathrm{i}Z_{\gamma_k}'\pi_{\gamma_k}^\circ\big)\big\}\big|\Sigma_{\gamma_k},\gamma_k,\mathcal{F}_t\big]\nonumber\\
&=& \exp\bigg\{\mathrm{tr}\Big(\mathrm{i}Z_{\gamma_k}'\pi_{0,\gamma_k|t}^\circ-\frac{1}{2}Z_{\gamma_k}'\Lambda_{0,\gamma_k|t}Z_{\gamma_k}\Sigma_{\gamma_k}\Big)\bigg\},
\end{eqnarray}
where $\mathrm{i}=\sqrt{-1}$ is the imaginary unit, $Z_{\gamma_k}$ is an $(n\times d)$ matrix, corresponding to the regime $\gamma_k$. Consequently, by the iterated expectation formula, conditional density function \eqref{08068}, and the above characteristic function of the matrix normal distribution, conditional on the regime $\gamma_k$ and information $\mathcal{F}_t$, a characteristic function of the random coefficient matrix $\pi_{\gamma_k}^\circ$ is obtained by
\begin{eqnarray}\label{08088}
\varphi(Z_{\gamma_k}|\gamma_k,\mathcal{F}_t)&=&\mathbb{E}\big[\exp\big\{\mathrm{tr}\big(\mathrm{i} Z_{\gamma_k}'\pi_{\gamma_k}^\circ\big)\big\}\big|\gamma_k,\mathcal{F}_t\big]\nonumber\\
&=&\frac{\exp\big\{\mathrm{tr}\big(\mathrm{i} Z_{\gamma_k}'\pi_{0,\gamma_k|t}^\circ\big)\big\}\big|B_{t,\gamma_k}+V_{0,\gamma_k}\big|^{\nu_{0,\gamma_k|t}}|Z_{\gamma_k}'\Lambda_{0,\gamma_k|t}Z_{\gamma_k}|^{\nu_{0,\gamma_k|t}}}{\Gamma_{n}\big(\nu_{0,\gamma_k|t}/2\big)2^{n\nu_{0,\gamma_k|t}}}\nonumber\\
&\times& \mathcal{B}_{\nu_{0,\gamma_k|t}}\big(Z_{\gamma_k}'\Lambda_{0,\gamma_k|t}Z_{\gamma_k}\big(B_{t,\gamma_k}+V_{0,\gamma_k}\big)/4\big).
\end{eqnarray}
Similarly, it can be shown that
\begin{eqnarray}\label{08089}
\varphi(Z_{\delta_k}|\delta_k,\mathcal{F}_0)&=&\mathbb{E}\big[\exp\big\{\mathrm{tr}\big(\mathrm{i}Z_{\delta_k}'\pi_{\delta_k}^\circ\big)\big\}\big|\delta_k,\mathcal{F}_0\big]\nonumber\\
&=&\frac{\exp\big\{\mathrm{tr}\big(\mathrm{i}Z_{\delta_k}'\pi_{0,\delta_k}^\circ\big)\big\}\big|V_{0,\delta_k}\big|^{\nu_{0,\delta_k}}|Z_{\delta_k}'\Lambda_{0,\delta_k}Z_{\delta_k}|^{\nu_{0,\delta_k}}}{\Gamma_{n}(\nu_{0,\delta_k}/2)2^{n\nu_{0,\delta_k}}}\nonumber\\
&\times& \mathcal{B}_{\nu_{0,\delta_k}}\big(Z_{\delta_k}'\Lambda_{0,\delta_k}Z_{\delta_k}V_{0,\delta_k}/4\big).
\end{eqnarray}
The above characteristic functions can be used to obtain raw moments of the random coefficient matrix $\pi_{\beta}^\circ$ for given the regime vector $s$ and information $\mathcal{F}_t$. For example, since conditional on $s$ and $\mathcal{F}_t$, for $k=1,\dots,r_\gamma$ and $\ell=1,\dots,r_\delta$, coefficient matrices $\pi_{\beta_k}$ and $\pi_{\delta_\ell}$ are independent, we have that 
\begin{eqnarray}\label{08090}
&&\mathbb{E}\Big[\big(\pi_{\gamma_1}^\circ\big)_{i_{\gamma_1},j_{\gamma_1}}^{m_{\gamma_1}}\dots \big(\pi_{\gamma_{r_{\gamma}}}^\circ\big)_{i_{\gamma_{r_{\gamma}}},j_{\gamma_{r_{\gamma}}}}^{m_{\gamma_{r_{\gamma}}}}\big(\pi_{\delta_1}^\circ\big)_{i_{\delta_1},j_{\delta_1}}^{m_{\delta_1}}\dots \big(\pi_{\delta_{r_{\delta}}}^\circ\big)_{i_{\delta_{r_{\delta}}},j_{\delta_{r_{\delta}}}}^{m_{\delta_{r_{\delta}}}}\Big|s,\mathcal{F}_t\Big]\nonumber\\
&&=\prod_{k=1}^{r_\gamma}\frac{1}{\mathrm{i}^{m_{\gamma_k}}}\frac{\partial^{m_{\gamma_k}}\varphi(Z_{\gamma_k}|\gamma_k,\mathcal{F}_t)}{\partial (Z_{\gamma_k})_{i_{\gamma_k},j_{\gamma_k}}^{m_{\gamma_k}}}\times\prod_{\ell=1}^{r_\delta}\frac{1}{\mathrm{i}^{m_{\delta_\ell}}}\frac{\partial^{m_{\delta_\ell}}\varphi(Z_{\delta_\ell}|\delta_\ell,\mathcal{F}_0)}{\partial (Z_{\gamma_k})_{i_{\delta_\ell},j_{\delta_\ell}}^{m_{\gamma_k}}},
\end{eqnarray}
where for a generic $(n\times m)$ matrix $O$, $(O)_{i,j}$ denotes an $(i,j)$--th element of the matrix $O$ for $i=1,\dots,n$ and $j=1,\dots,m$. The partial derivatives can be calculated by the numerical methods. The raw moments may be used to obtain forecast of the vector of endogenous variables. In particular, conditional on $\bar{s}_{t+2}$, the optimal forecast, which minimizes the mean squared errors for forecast horizon 2 at forecast origin $t$ equals an expectation of the vector of endogenous variables at time $(t+2)$ for given $\mathcal{F}_t$. Thus, the forecast is given by the following equation
\begin{eqnarray}\label{08091}
\mathbb{E}\big[y_{t+2}\big|\bar{s}_{t+2},\mathcal{F}_t\big]&=&\mathbb{E}\big[A_{0,s_{t+2}}\big|\bar{s}_{t+2},\mathcal{F}_t\big]+\mathbb{E}\big[A_{1,s_{t+2}}A_{0,s_{t+1}}\big|\bar{s}_{t+2},\mathcal{F}_t\big]\psi_{t+1}\\
&+&\sum_{k=1}^p\mathbb{E}\big[A_{1,s_{t+2}}A_{k,s_{t+1}}|\bar{s}_{t+2},\mathcal{F}_t\big]y_{t+1-k}+\sum_{k=2}^p\mathbb{E}\big[A_{k,s_{t+2}}|\bar{s}_{t+2},\mathcal{F}_t\big]y_{t+2-k},\nonumber
\end{eqnarray}
where the conditional expectations $\mathbb{E}\big[A_{k,s_{t+2}}\big|\bar{s}_{t+2},\mathcal{F}_t\big]$ for $k=0,2,\dots,p$ are calculated by equations \eqref{08069} and \eqref{08070} and the conditional expectations $\mathbb{E}\big[A_{1,s_{t+2}}A_{k,s_{t+1}}\big|\bar{s}_{t+2},\mathcal{F}_t\big]$ for $k=0,\dots,p$ are calculated equation \eqref{08090}. To illustrative purpose, we assume that $s_{t+1},s_{t+2}\in \mathcal{A}_{\bar{s}_t}\cap \mathcal{A}_{\bar{s}_t^c}$ and positions of the regimes $s_{t+1}$ and $s_{t+2}$ in the regime vector $\gamma$ are $k_{1*}$ and $k_{2*}$, respectively, that is,
\begin{equation}\label{08092}
k_{i*}=\{k\in\{1,\dots,r_\gamma\}|\gamma_k=s_{t+i},k=1,\dots,r_\gamma\}
\end{equation}
for $i=1,2$. Then, we have that for $i=1,\dots,n$, $j=1,\dots,l$, and $k=0$,
\begin{equation}\label{08093}
\Big(\mathbb{E}\big[A_{1,s_{t+2}}A_{k,s_{t+1}}|\bar{s}_{t+2},\mathcal{F}_t\big]\Big)_{i,j}=- \sum_{\ell=1}^n\frac{\partial \varphi(Z_{\gamma_{k_{2*}}}|\gamma_{k_{2*}},\mathcal{F}_t)}{\partial (Z_{\gamma_{k_{2*}}})_{i,l+\ell}}\frac{\partial \varphi(Z_{\gamma_{k_{1*}}}|\gamma_{k_{1*}},\mathcal{F}_t)}{\partial (Z_{\gamma_{k_{1*}}})_{\ell,j}},
\end{equation}
for $s_{t+1}=s_{t+2}$, $i=1,\dots,n$, and $k=1$,
\begin{eqnarray}\label{08094}
&&\Big(\mathbb{E}\big[A_{1,s_{t+2}}A_{k,s_{t+1}}|\bar{s}_{t+2},\mathcal{F}_t\big]\Big)_{i,i}\nonumber\\
&&=- \sum_{\ell=1,\ell\neq i}^n\frac{\partial \varphi(Z_{\gamma_{k_{1*}}}|\gamma_{k_{1*}},\mathcal{F}_t)}{\partial (Z_{\gamma_{k_{1*}}})_{i,l+\ell}}\frac{\partial \varphi(Z_{\gamma_{k_{1*}}}|\gamma_{k_{1*}},\mathcal{F}_t)}{\partial (Z_{\gamma_{k_{1*}}})_{l+\ell,i}}-\frac{\partial^2 \varphi(Z_{\gamma_{k_{1*}}}|\gamma_{k_{1*}},\mathcal{F}_t)}{\partial (Z_{\gamma_{k_{1*}}})_{i,l+i}^2},
\end{eqnarray}
and for other cases ($i,j=1,\dots,n$ and $k=1,\dots,p$),
\begin{equation}\label{08095}
\Big(\mathbb{E}\big[A_{1,s_{t+2}}A_{k,s_{t+1}}|\bar{s}_{t+2},\mathcal{F}_t\big]\Big)_{i,j}=-\sum_{\ell=1}^n\frac{\partial \varphi(Z_{\gamma_{k_{2*}}}|\gamma_{k_{2*}},\mathcal{F}_t)}{\partial (Z_{\gamma_{k_{2*}}})_{i,l+\ell}}\frac{\partial \varphi(Z_{\gamma_{k_{1*}}}|\gamma_{k_{1*}},\mathcal{F}_t)}{\partial (Z_{\gamma_{k_{1*}}})_{l+(k-1)p+\ell,j}}.
\end{equation}
Because the exact calculation of forecast of the process of endogenous variables is complicated, we consider an approximation, which is used to calculate the forecast of the endogenous variables in \citeA{Banbura10}. For $u=t+1,\dots,T$, by the iterated expectation formula, conditional on the regime vector $s$, the exact forecast is given by the following equation
\begin{equation}\label{08157}
\mathbb{E}\big[y_u\big|s,\mathcal{F}_t\big]=\mathbb{E}\big[\Pi_{s_u}\mathsf{Y}_u\big|s,\mathcal{F}_t\big]=\mathbb{E}\big[\mathbb{E}\big[\Pi_{s_u}\big|s,\mathcal{F}_{u-1}\big]\mathsf{Y}_u\big|s,\mathcal{F}_t\big].
\end{equation}
\citeA{Banbura10} approximate the last expression by $\mathbb{E}\big[\Pi_{s_u}\big|s,\mathcal{F}_{u-1}\big]\mathbb{E}\big[\mathsf{Y}_u\big|s,\mathcal{F}_t\big].$ Consequently, the forecast is approximated by
\begin{equation}\label{08158}
\mathbb{E}\big[y_u\big|s,\mathcal{F}_t\big]\approx\mathbb{E}\big[\Pi_{s_u}\big|s,\mathcal{F}_{u-1}\big]\mathbb{E}\big[\mathsf{Y}_u\big|s,\mathcal{F}_t\big].
\end{equation}
For $t=u-1$, the approximation becomes exact, namely,
\begin{equation}\label{08159}
\mathbb{E}\big[y_u\big|s,\mathcal{F}_{u-1}\big]=\mathbb{E}\big[\Pi_{s_u}\big|s,\mathcal{F}_{u-1}\big]\mathsf{Y}_u.
\end{equation}
However, for $u=t+2,\dots,T$, one should study the quality of the simple approximation.

\subsubsection{Minnesota Prior}

In practice, one usually adopts Minnesota prior to estimating the parameters of the VAR$(p)$ process. The first version of  Minnesota prior was introduced by \citeA{Litterman79}. Also, \citeA{Banbura10} used Minnesota prior for large Bayesian VAR and showed that the forecast of large Bayesian VAR is better than small Bayesian VAR. However, there are many different variants of the Minnesota prior, we consider a prior, which is included in \citeA{Miranda18}. The idea of Minnesota prior is that it shrinks diagonal elements of the matrix $A_{1,s_t}$ toward $\phi_i$ and off--diagonal elements of $A_{1,s_t}$ and all elements of other matrices $A_{0,s_t},A_{2,s_t},\dots,A_{p,s_t}$ toward 0, where $\phi_i$ is 0 for a stationary variable $y_{i,t}$ and 1 for a variable with unit root $y_{i,t}$.  For the prior, it is assumed that conditional on $\Sigma_{s_t}$, $s_t$, and $\mathcal{F}_0$, $A_{0,s_t},A_{1,s_t},\dots,A_{p,s_t}$ are jointly normally distributed, and for $(i,j)$--th element of the matrix $A_{\ell,s_t}$ $(\ell=0,\dots,p)$, it holds
\begin{equation}\label{08096}
\mathbb{E}\big((A_{\ell,s_t})_{i,j}\big|\Sigma_{s_t},s_t,\mathcal{F}_0\big)=\begin{cases}
\phi_i & \text{if}~~~i=j,~\ell=1,\\
0 & \text{if}~~~\text{otherwise}
\end{cases},
\end{equation}
\begin{equation}\label{08097}
\text{Var}\big((A_{0,s_t})_{i,j}\big|\Sigma_{s_t},s_t,\mathcal{F}_0\big)=(\sigma_{i,s_t}/\varepsilon_{s_t})^2,
\end{equation}
and
\begin{equation}\label{08098}
\text{Var}\big((A_{\ell,s_t})_{i,j}\big|\Sigma_{s_t},s_t,\mathcal{F}_0\big)=\begin{cases}
\displaystyle \bigg(\frac{\sigma_{i,s_t}}{\ell^{\lambda_{2,s_t}}\lambda_{1,s_t}\tau_{i,s_t}}\bigg)^2 & \text{if}~~~i=j,\\
\displaystyle \bigg(\frac{\sigma_{i,s_t}}{\ell^{\lambda_{2,s_t}}\lambda_{1,s_t}\tau_{j,s_t}}\bigg)^2 & \text{if}~~~\text{otherwise}
\end{cases}~~~\text{for}~\ell=1,\dots,p,
\end{equation}
where $\sigma_{i,s_t}^2$ is an $(i,i)$--th element of the random covariance matrix $\Sigma_{s_t}$. The parameter $\varepsilon_{s_t}^2$ is a small number and it corresponds to an uninformative diffuse prior for $(A_{0,s_t})_{i,j}$, the parameter $\lambda_{1,s_t}$ controls the overall tightness of the prior distribution, the parameter $\lambda_{2,s_t}$ controls amount of information prior information at higher lags, and $\tau_{i,s_t}$ is a scaling parameter, see \citeA{Miranda18}. Thus, the factor $1/\ell^{2\lambda_{2,s_t}}$ represents a rate at which prior variance decreases with increasing lag length. 

According to \citeA{Banbura10}, it can be shown that the following equation satisfies the prior conditions \eqref{08096}, \eqref{08097}, and \eqref{08098}
\begin{equation}\label{08099}
\hat{y}_{s_t}^\circ=\Pi_{s_t}\hat{\mathsf{Y}}_{s_t}^\circ+\hat{\xi}_{s_t}^\circ,
\end{equation}
where $\hat{y}_{s_t}^\circ$ and $\hat{\mathsf{Y}}_{s_t}^\circ$ are $(n\times d)$ and $(d\times d)$ matrices of dummy variables and are defined by
\begin{equation}\label{08100}
\hat{y}_{s_t}^\circ:=\big[0_{[n\times l]}:\lambda_{1,s_t}\text{diag}\{\phi_1\tau_{1,s_t},\dots,\phi_n\tau_{n,s_t}\}:0_{[n\times n(p-1)]}\big]
\end{equation}
and
\begin{equation}\label{08101}
\hat{\mathsf{Y}}_{s_t}^\circ:=\begin{bmatrix}
\varepsilon_{s_t} I_l & 0_{[l\times np]}\\
0_{[np\times l]} & \lambda_{1,s_t}\big(J_{s_t}\otimes\text{diag}\{\tau_{1,s_t},\dots,\tau_{n,s_t}\}\big)
\end{bmatrix}
\end{equation}
with $J_{s_t}:=\text{diag}\{1^{\lambda_{2,s_t}},\dots,p^{\lambda_{2,s_t}}\}$, respectively, and $\hat{\xi}_{s_t}^\circ:=[\xi_{1,s_t}:\dots:\xi_{d,s_t}]$ is an $(n\times d)$ matrix, whose columns are conditional independent for the given covariance matrix $\Sigma_{s_t}$, regime $s_t$, and initial information $\mathcal{F}_0$ and for $i=1,\dots,d$, each column has a distribution $\xi_{i,s_t}|\Sigma_{s_t},s_t,\mathcal{F}_0\sim \mathcal{N}(0,\Sigma_{s_t})$. Note that one can add constraints for elements of the coefficient matrix $\Pi_{s_t}$ to the matrices of dummy variables. It is worth mentioning that the matrices of dummy variables $\hat{y}_{s_t}$ and $\hat{\mathsf{Y}}_{s_t}$ should not depend on the covariance matrix $\Sigma_{s_t}$. If the dummy variables depend on the covariance matrix, an OLS estimator $\hat{\pi}_{s_t}$, and matrix $\Lambda_{0,s_t}$ depend on the covariance matrix $\Sigma_{s_t}$, see below. Consequently, in this case, one can not use the results of Proposition \ref{prop02}. For this reason, we choose the prior condition \eqref{08098}, c.f. \citeA{Banbura10}. Equation \eqref{08099}, can be written by
\begin{equation}\label{08102}
\hat{y}_{s_t}=\big((\hat{\mathsf{Y}}_{s_t}^\circ)'\otimes I_n\big)\pi_{s_t}+\hat{\xi}_{s_t},
\end{equation}
where $\hat{y}_{s_t}$ and $\hat{\xi}_{s_t}$ are $([nd]\times 1)$ vectors and are vectorizations of the matrix of dummy variables $\hat{y}_{s_t}^\circ$ and matrix of the residual process $\hat{\xi}_{s_t}^\circ$, respectively, i.e., $\hat{y}_{s_t}:=\text{vec}(\hat{y}_{s_t}^\circ)$ and $\hat{\xi}_{s_t}:=\text{vec}(\hat{\xi}_{s_t}^\circ)$. It follows from equation \eqref{08102} that
\begin{equation}\label{08103}
\pi_{s_t}\overset{d}{=}\big(((\hat{\mathsf{Y}}_{s_t}^\circ (\hat{\mathsf{Y}}_{s_t}^\circ)')^{-1}\hat{\mathsf{Y}}_{s_t}^\circ)\otimes I_n\big)\hat{y}_{s_t}+\big(((\hat{\mathsf{Y}}_{s_t}^\circ (\hat{\mathsf{Y}}_{s_t}^\circ)')^{-1}\hat{\mathsf{Y}}_{s_t}^\circ)\otimes I_n\big)\hat{\xi}_{s_t},
\end{equation}
where $d$ denotes equal distribution. It should be noted that the first term of the right--hand side of the above equation is a vecorization of the ordinary least square (OLS) estimator of the coefficient matrix $\Pi_{s_t}$, namely, $\pi_{0,s_t}:=\text{vec}\big(\hat{\Pi}_{s_t}\big)=\text{vec}\big(\hat{y}_{s_t}^\circ(\hat{\mathsf{Y}}_{s_t}^\circ)'(\hat{\mathsf{Y}}_{s_t}^\circ(\hat{\mathsf{Y}}_{s_t}^\circ)')^{-1}\big)$. Consequently, since $\hat{\xi}_{s_t}|\Sigma_{s_t},s_t,\mathcal{F}_0\sim \mathcal{N}\big(0,I_d\otimes \Sigma_{s_t}\big)$, conditional on $\Sigma_{s_t}$, $s_t$, and $\mathcal{F}_0$, a distribution of the random coefficient vector $\pi_{s_t}$ is given by
\begin{equation}\label{08104}
\pi_{s_t}~|~\Sigma_{s_t},s_t,\mathcal{F}_0\sim \mathcal{N}\Big(\pi_{0,s_t},\big(\Lambda_{0,s_t}\otimes \Sigma_{s_t}\big)\Big),
\end{equation}
where $\Lambda_{0,s_t}:=(\hat{\mathsf{Y}}_{s_t}^\circ(\hat{\mathsf{Y}}_{s_t}^\circ)')^{-1}$ is a $(d\times d)$ diagonal matrix. Therefore, conditional on $\Sigma_{s_t}$, $s_t$, and $\mathcal{F}_0$, columns of the random coefficient matrix $\Pi_{s_t}$ are independent. For the moment conditions \eqref{08096}, \eqref{08097}, and \eqref{08098}, we can use the results of Proposition \ref{prop02}.

\subsection{Simulation Methods}

By applying Propositions \ref{prop02} and \ref{prop03}, one can obtain exact density function of the random vector $\bar{y}_t$, namely,
\begin{eqnarray}\label{08105}
f(\bar{y}_t|\mathcal{F}_0)=\sum_{\bar{s}_t}f(\bar{y}_t|\bar{s}_t,\mathcal{F}_0)\prod_{i=0}^N\frac{\Gamma\big(\sum_{j=1}^N\alpha_{ij}\big)}{\prod_{j=1}^N\Gamma(\alpha_{ij})}\frac{\prod_{j=1}^N\Gamma(\alpha_{ij}+n_{ij}(\bar{s}_t))}{\Gamma\big(\sum_{j=1}^N(\alpha_{ij}+n_{ij}(\bar{s}_t))\big)}.
\end{eqnarray}
However, this exact density function has the following two main disadvantages:
\begin{itemize}
\item[$(i)$] it is difficult to obtain characteristics (such as mean, quantile, marginal densities, distribution, and so on) of the mixture density function $f(\bar{y}_t|\mathcal{F}_0)$,
\item[$(ii)$] and it is difficult to calculate the sum with respect to $\bar{s}_t$. For example, if the length of the regime vector $\bar{s}_t$ equals 30 and the regime number equals 3, then we have to calculate $3^{30}\approx 2.06\times 10^{14}$ summands.
\end{itemize}
If the dimensions increase, the disadvantages are seriously worsen. Therefore, from a practical point of view, we need to develop Monte--Carlo simulation method. 

\subsubsection{General Method}

According to the conditional probability formula and Proposition \ref{prop01}, we have that
\begin{eqnarray}\label{08106}
f(\bar{y}_t^c,\pi_{\hat{s}},\Sigma_{\hat{s}},s,\mathsf{P}|\mathcal{F}_t)&=&f(\bar{y}_t^c|\pi_{\beta},\Sigma_{\beta},\bar{s}_t^c,\mathcal{F}_t)f_*(\pi_{\delta}|\Sigma_{\delta},\delta,\mathcal{F}_0)f_*(\Sigma_{\delta}|\delta,\mathcal{F}_0)\\
&\times& f(\bar{s}_t^c|s_t,\mathsf{P},\mathcal{F}_0)f(\pi_{\alpha},\Sigma_{\alpha},\bar{s}_t,\mathsf{P}|\mathcal{F}_t).\nonumber
\end{eqnarray}
The above equation tells us that how to generate random samples $(\bar{y}_t^c,\pi_{\hat{s}_t},\Sigma_{\hat{s}_t},s,\mathsf{P})$ for given information $\mathcal{F}_t$. The direction of our simulation method move toward from right to left for the above equation. To generate the random samples, firstly, we generate the random coefficient vector, random covariance matrix, regime vector, and transition probability matrix $(\pi_\alpha,\Sigma_\alpha,\bar{s}_t,\mathsf{P})$ from the posterior density function $f(\pi_\alpha,\Sigma_\alpha,\bar{s}_t,\mathsf{P}|\mathcal{F}_t)$. Next, using the regime vector $\bar{s}_t$ and transition probability matrix $\mathsf{P}$, generate regime vector $\bar{s}_t^c$ from the conditional density function $f(\bar{s}_t^c|s_t,\mathsf{P},\mathcal{F}_0)$, so on.

First, we consider a simulation method that generate the random coefficient vector, random covariance matrix, regime vector, and transition probability matrix $(\pi_\alpha,\Sigma_\alpha,\bar{s}_t,\mathsf{P})$ from the posterior density function $f(\pi_\alpha,\Sigma_\alpha,\bar{s}_t,\mathsf{P}|\mathcal{F}_t)$. We develop the Gibbs sampling method to generate $(\pi_\alpha,\Sigma_\alpha,\bar{s}_t,\mathsf{P})$. In the Bayesian statistics, the Gibbs sampling is often used when the joint distribution is not known explicitly or is difficult to sample from directly, but the conditional distribution of each variable is known and is easy to sample from. According to Proposition \ref{prop01}, constructing the Gibbs sampler to approximate the joint posterior distribution $f(\bar{s}_t,\mathsf{P}|\mathcal{F}_t)$ is straightforward: set initial values $\big(\pi_{\alpha(0)}(0),\Sigma_{\alpha(0)}(0),\bar{s}_t(0),\mathsf{P}(0)\big)$ and new values $\big(\pi_{\alpha(\ell)}(\ell),\Sigma_{\alpha(\ell)}(\ell),\bar{s}_t(\ell),\mathsf{P}(\ell)\big)$, $\ell=1,\dots,\mathcal{L}$ can be generated by
\begin{itemize}
\item generate $\bar{s}_t(\ell+1)$ from $f\big(\bar{s}_t\big|\pi_{\alpha(\ell)}(\ell),\Sigma_{\alpha(\ell)}(\ell),\mathsf{P}(\ell),\mathcal{F}_t\big)$,
\item generate $\Sigma_{\alpha(\ell+1)}(\ell+1)$ from $f\big(\Sigma_{\alpha}\big|\alpha(\ell+1),\mathcal{F}_t\big)$,
\item generate $\pi_{\alpha(\ell+1)}(\ell+1)$ from $f\big(\pi_{\alpha}\big|\Sigma_{\alpha(\ell+1)}(\ell+1),\alpha(\ell+1),\mathcal{F}_t\big)$,
\item generate $\mathsf{P}(\ell+1)$ from $f\left(\mathsf{P}|\bar{s}_t(\ell+1),\mathcal{F}_0\right)$,
\end{itemize}
where $\alpha(\ell+1)$ is the duplication removed regime vector, corresponding to the regime vector $\bar{s}_t(\ell+1)$.

Now, we consider a sampling method that generate $\bar{s}_t(\ell+1)$ from $f\big(\bar{s}_t\big|\lambda_t(\ell),\mathcal{F}_t\big)$, where the parameter vector at iteration $\ell$, $\lambda_t(\ell):=\big(\pi_{\alpha(\ell)}(\ell),\Sigma_{\alpha(\ell)}(\ell),\mathsf{P}(\ell)\big)$ corresponds to the regime vector $\bar{s}_t$. Here we follow some results of the book of \citeA{Hamilton94}, see also  \citeA{Battulga23b} and \citeA{Battulga22b}. If we assume that the regime--switching process in regime $j$ at time $u$, then the conditional density function of the random vector $y_u$ is given by the following equation
\begin{eqnarray}\label{08108}
\eta_{u,j}(\ell)&:=&f(y_u|\lambda_t(\ell),s_u(\ell)=j,\mathcal{F}_{u-1})\\
&=&\frac{1}{(2\pi)^{n/2}|\Sigma_j(\ell)|^{1/2}}\exp\bigg\{-\frac{1}{2}\Big(y_u-(\mathsf{Y}_u'\otimes I_n)\pi_j(\ell)\Big)'\Sigma_j^{-1}(\ell)\Big(y_u-(\mathsf{Y}_u'\otimes I_n)\pi_j(\ell)\Big)\bigg\},\nonumber
\end{eqnarray}
for $u=1,\dots,t$ and $j=1,\dots,N$. For all $u=1,\dots,t$, we collect the conditional density functions of $y_u$ into an $(N\times 1)$ vector $\eta_t$, that is, $\eta_u(\ell):=(\eta_{u,1}(\ell),\dots,\eta_{u,N}(\ell))'$. Let us denote a probabilistic inference about the value of the regime--switching process $s_u$ is equal to $j$, based on the random coefficient vector $\pi_{\alpha(\ell)}(\ell)$, random covariance matrix $\Sigma_{\alpha(\ell)}(\ell)$, information $\mathcal{F}_u$, and transition probability matrix $\mathsf{P}(\ell)$ by $\mathbb{P}(s_u=j|\lambda_t(\ell),\mathcal{F}_u)$. Collect these conditional probabilities $\mathbb{P}(s_u=j|\lambda_t(\ell),\mathcal{F}_u)$ for $j=1,\dots,N$ into an $(N\times 1)$ vector $z_{u|u}(\ell)$, that is, $z_{u|u}(\ell):=\big(\mathbb{P}(s_u=1|\lambda_t(\ell),\mathcal{F}_u),\dots,\mathbb{P}(s_u=N|\lambda_t(\ell),\mathcal{F}_u)\big)'$. Also, we need a probabilistic forecast about the value of the regime--switching process at time $u+1$ is equal to $j$ conditional on the random coefficient vector $\pi_{\alpha(\ell)}(\ell)$, random covariance matrix $\Sigma_{\alpha(\ell)}(\ell)$, transition probability matrix $\mathsf{P}(\ell)$, and data up to and including time $u$, $\mathcal{F}_u$. Collect these forecasts into an $(N\times 1)$ vector $z_{u+1|u}(\ell)$, that is, $z_{u+1|u}(\ell):=\big(\mathbb{P}(s_{u+1}=1|\lambda_t(\ell),\mathcal{F}_u),\dots,\mathbb{P}(s_{u+1}=N|\lambda_t(\ell),\mathcal{F}_u)\big)'$.  

The probabilistic inference and forecast for each time $u=1,\dots,t$ can be found by iterating on the following pair of equations: 
\begin{equation}\label{08109}
z_{u|u}(\ell)=\frac{(z_{u|u-1}(\ell)\odot\eta_u(\ell))}{i_N'(z_{u|u-1}(\ell)\odot\eta_u(\ell))}~~~\text{and}~~~z_{u+1|u}(\ell)=\hat{\mathsf{P}}(\ell)'z_{u|u}(\ell),~~~u=1,\dots,t,
\end{equation}
where $\odot$ is the Hadamard product of two vectors, $\eta_u(\ell)$ is the $(N\times 1)$ vector, whose $j$--th element is given by equation \eqref{08108}, $\hat{\mathsf{P}}(\ell)$ is the $(N\times N)$ transition probability matrix, which omits the first row of the transition probability matrix $\mathsf{P}(\ell)$, and $i_N$ is an $(N\times 1)$ vector, whose elements equal 1. Given a starting value $z_{1|0}(\ell):=\mathsf{P}_0(\ell)'$ one can iterate on \eqref{08109} for $u=1,\dots,t$ to calculate the values of $z_{u|u}(\ell)$ and $z_{u+1|u}(\ell)$. 

To obtain marginal distributions of the regime vector $\bar{s}_t$ conditional on the transition probability matrix $\mathsf{P}(\ell)$ and information $\mathcal{F}_t$, let us introduce $(N\times 1)$ smoothed inference vector $z_{u|t}(\ell):=\big(\mathbb{P}(s_u=1|\lambda_t(\ell),\mathcal{F}_t),\dots,\mathbb{P}(s_u=N|\lambda_t(\ell),\mathcal{F}_t)\big)'$ for $u=1,\dots,t$. In practice, a popular method to calculate smoothed probability inference is the \citeA{Kim94}'s smoothing algorithm, which is based on approximation. In this paper, we introduce a new and simple smoothing method, which is also based on an approximation and is given the following Proposition.

\begin{proposition}\label{prop04}
Let us assume that 
\begin{equation}\label{08163}
\lim_{t\to\infty}f_*\big(\pi_{s_{t+1}}(\ell),\Sigma_{s_{t+1}}(\ell)\big|\bar{s}_{t+1},\lambda_t(\ell),\mathcal{F}_t\big)=1,
\end{equation}
where the density function is defined by
\begin{equation}\label{08164}
f_*\big(\pi_{s_{t+1}}(\ell),\Sigma_{s_{t+1}}(\ell)\big|\bar{s}_{t+1},\lambda_t(\ell),\mathcal{F}_t\big)=\begin{cases}
1 & \mathrm{if}~~~s_{t+1}\in \mathcal{A}_{\bar{s}_t},\\
f\big(\pi_{s_{t+1}}(\ell),\Sigma_{s_{t+1}}(\ell)\big|\mathcal{F}_0\big) & \mathrm{if}~~~s_{t+1}\not\in \mathcal{A}_{\bar{s}_t}.
\end{cases}
\end{equation}
Then for sufficiently large $t$, the smoothed probability inference vectors are approximated by
\begin{equation}\label{08110}
z_{t-1|t}(\ell)=\frac{1}{i_N'(z_{t|t-1}(\ell)\odot\eta_t(\ell))}\big(\hat{\mathsf{P}}(\ell)\mathsf{H}_t(\ell)i_N\big)\odot z_{t-1|t-1}(\ell)
\end{equation}
and for $u=t-2,\dots,1$,
\begin{equation}\label{08180}
z_{u|t}(\ell)=\frac{1}{i_N'(z_{u+1|u}(\ell)\odot\eta_{u+1}(\ell))}\Big(\hat{\mathsf{P}}(\ell)\mathsf{H}_{u+1}(\ell)\big(z_{u+1|t}(\ell)\oslash z_{u+1|u+1}(\ell)\big)\Big)\odot z_{u|u}(\ell),
\end{equation}
where $\oslash$ is an element--wise division of two vectors and $\mathsf{H}_{u+1}(\ell):=\mathrm{diag}\{\eta_{u+1,1}(\ell),\dots,\eta_{u+1,N}(\ell)\}$ is an $(N\times N)$ diagonal matrix. For $u=2,\dots,t$, joint density function of the regimes $s_{u-1}$ and $s_u$ is approximated by
\begin{equation}\label{08165}
f(s_{u-1},s_u|\lambda_t(\ell),\mathcal{F}_t)=\frac{\big(z_{u|t}(\ell)\big)_{s_u}\eta_{u,s_u}(\ell)p_{s_{u-1}s_u}\big(z_{u-1|u-1}(\ell)\big)_{s_{u-1}}}{\big(z_{u|u}(\ell)\big)_{s_u}i_N'(z_{u|u-1}\odot\eta_u)},
\end{equation}
where for a generic vector $o$, $(o)_j$ denotes $j$--th element of the vector $o$.
\end{proposition}

The smoothed probabilities $z_{u|t}(\ell)$ are found by iterating on \eqref{08110} and \eqref{08180} backward for $u=t-1,\dots,1$. This iteration uses probabilistic inferences and forecasts $z_{u|u}(\ell)$ and $z_{u+1|u}(\ell)$ for $u=1,\dots,t-1$, which are obtained from \eqref{08109}. Thus, the simulation method that generate $\bar{s}_t(\ell)$ for given $\mathsf{P}(\ell)$ and $\mathcal{F}_t$ as follows:
\begin{itemize}
\item generate $s_u(\ell+1)$ from $z_{u|t}(\ell)$ for $u=1,\dots,t.$
\end{itemize}
Collect $s_u(\ell+1)$ for $u=1,\dots,t$ into $(t\times 1)$ vector $\bar{s}_t(\ell+1)$, namely, $\bar{s}_t(\ell+1):=(s_1(\ell+1),\dots,s_t(\ell+1))'$. The smoothing method is not only used to Bayesian estimation but also maximum likelihood estimation (ML) of parameters of a model with regime switch. Also, to estimate parameters of the model,  one can use the joint density function. It is worth mentioning that if vector of the endogenous variables $y_t$ does not depend $(\pi_{\alpha},\Sigma_\alpha)$, then the smoothing method and joint density function become exact. Therefore, when they are used to the ML estimation of the model's parameters, the ML estimation becomes exact.  

Let us consider a simulation method that generate the random coefficient vector $\pi_{\alpha(\ell+1)}(\ell+1)$ and random covariance matrix $\Sigma_{\alpha(\ell+1)}(\ell+1)$ from the density function $f\big(\pi_{\alpha(\ell+1)},\Sigma_{\alpha(\ell+1)}\big|\alpha(\ell+1),\mathcal{F}_t\big)$. Let $\mathcal{A}_{{\bar{s}}_t(\ell+1)}=\big\{\alpha_1(\ell+1),\dots,\alpha_{r_{\alpha(\ell+1)}}(\ell+1)\big\}$ be the duplication removed set, corresponding the regime vector $\bar{s}_t(\ell+1)$. Then, according to equations \eqref{08010}, \eqref{08067} and \eqref{08068}, a simulation method that generates $\big(\pi_{\gamma_k(\ell+1)}(\ell),\Sigma_{\gamma_k(\ell+1)}(\ell+1)\big)$ as follows: for $k=1,\dots,r_{\alpha(\ell+1)}$,  
\begin{itemize}
\item generate $\Sigma_{\alpha_k(\ell+1)}(\ell+1)$ from $\mathcal{IW}\Big(\nu_{0,\alpha_k(\ell+1)|t},B_{t,\alpha_k(\ell+1)}+V_{0,\alpha_k(\ell+1)}\Big),$
\item generate $\pi_{\alpha_k(\ell+1)}(\ell+1)$ from $\mathcal{N}\Big(\pi_{0,\alpha_k(\ell+1)|t},A_{\alpha_k(\ell+1)|t}\Big)$,
\end{itemize}
where $\nu_{0,\alpha_k(\ell+1)|t}:=\nu_{0,\alpha_k(\ell+1)}+q_{t,\alpha_k(\ell+1)}$ and $q_{t,\alpha_k(\ell+1)}$ is a number of regimes in the regime vector $\bar{s}_t(\ell+1)$ that equal the regime $\alpha_k(\ell+1)$. 

To generate $\mathsf{P}(\ell+1)$ from $f(\mathsf{P}|\bar{s}_t(\ell+1),\mathcal{F}_0)$, we apply equation \eqref{08017} in Proposition \ref{prop01}, density function \eqref{08074}, equation \eqref{08075} in Proposition \ref{prop03}, and the Bayesian formula. Then, we have that
\begin{equation}\label{08107}
f(\mathsf{P}|\bar{s}_t(\ell+1),\mathcal{F}_0)=\prod_{i=0}^N\frac{\Gamma\big(\sum_{j=1}^N(\alpha_{ij}+n_{ij}(\bar{s}_t(\ell+1)))\big)}{\prod_{j=1}^N\Gamma(\alpha_{ij}+n_{ij}(\bar{s}_t(\ell+1)))}\prod_{j=1}^Np_{ij}^{\alpha_{ij}+n_{ij}(\bar{s}_t(\ell+1))-1}.
\end{equation}
Thus, one can deduce that conditional on $\bar{s}_t(\ell+1)$ and $\mathcal{F}_0$, for $i=0,\dots,N$, each row of the transition probability matrix $\mathsf{P}$ are independent and has Dirichlet distribution with parameter $\alpha_i(\bar{s}_t(\ell+1)):=(\alpha_{i1}+n_{i1}(\bar{s}_t(\ell+1)),\dots,\alpha_{iN}+n_{iN}(\bar{s}_t(\ell+1)))'$. Consequently, a simulation method that generate $\mathsf{P}(\ell+1)$ for given $\bar{s}_t(\ell+1)$ and $\mathcal{F}_0$ as follows:
\begin{itemize}
\item generate $\mathsf{P}_i(\ell+1)$ from $\text{Dir}(\alpha_i(\bar{s}_t(\ell+1)))$ for $i=0,\dots,N.$ 
\end{itemize}
Collect $\mathsf{P}_i(\ell+1)$ for $i=0,\dots,N$ into an $([N+1]\times N)$ matrix $\mathsf{P}(\ell+1)$, that is, $\mathsf{P}(\ell+1):=[\mathsf{P}_0(\ell+1)':\dots:\mathsf{P}_N(\ell+1)']'.$ 

Second, for $\ell=1,\dots,\mathcal{L}$, we consider a simulation method that generates the regime vector $\bar{s}_t^c(\ell)$ from the density function $f(\bar{s}_t^c|s_t(\ell),\mathsf{P}(\ell),\mathcal{F}_0)$. Note that conditional on the transition probability matrix $\mathsf{P}(\ell)$ and initial information $\mathcal{F}_0$, the regime--switching process $s_t$ is a Markov chain. Thus, to sample the regime vector $\bar{s}_t^c(\ell)$ from the density function $f(\bar{s}_t^c|s_t(\ell),\mathsf{P}(\ell),\mathcal{F}_0)$, we can use the Markov property. That is, we have that
\begin{equation}\label{08111}
f(\bar{s}_t^c|s_t(\ell),\mathsf{P}(\ell),\mathcal{F}_0)=f(s_T|s_{T-1},\mathsf{P}(\ell),\mathcal{F}_0)\dots f(s_{t+1}|s_t(\ell),\mathsf{P}(\ell),\mathcal{F}_0). 
\end{equation}
Thus, a simulation method that generate $\bar{s}_t^c(\ell)$ for given $s_t(\ell)$, $\mathsf{P}(\ell)$, and $\mathcal{F}_0$ as follows:
\begin{itemize}
\item generate $s_{t+1}(\ell)$ from $f(s_{t+1}|s_t(\ell),\mathsf{P}(\ell),\mathcal{F}_0)$,
\item generate $s_{t+2}(\ell)$ from $f(s_{t+2}|s_{t+1}(\ell),\mathsf{P}(\ell),\mathcal{F}_0)$,
\item[] $\dots$
\item generate $s_{T}(\ell)$ from $f(s_{T}|s_{T-1}(\ell),\mathsf{P}(\ell),\mathcal{F}_0)$.
\end{itemize}
Collect $s_u(\ell)$ for $u=t+1,\dots,T$ into $([T-t]\times 1)$ vector $\bar{s}_t^c(\ell)$, namely, $\bar{s}_t^c(\ell):=(s_{t+1}(\ell),\dots,s_T(\ell))'$.

Third, for $\ell=1,\dots,\mathcal{L}$, we consider a simulation method that generates the coefficient vector $\pi_{\delta(\ell)}(\ell)$ and covariance matrix $\Sigma_{\delta(\ell)}(\ell)$ from the density function $f\big(\pi_{\delta(\ell)},\Sigma_{\delta(\ell)}\big|\delta(\ell),\mathcal{F}_0\big)$. Let $\mathcal{A}_{{\bar{s}}_t^c(\ell)}=\big\{\beta_1(\ell),\dots,\beta_{r_{\beta(\ell)}}(\ell)\big\}$ be the duplication removed set, corresponding the regime vector $\bar{s}_t^c(\ell)$. To eliminate unnecessary simulations, instead of the regimes in the set $\mathcal{A}_{{\bar{s}}_t^c(\ell)}$, one should consider regimes in difference set of the sets $\mathcal{A}_{{\bar{s}}_t(\ell)}$ and $\mathcal{A}_{{\bar{s}}_t^c(\ell)}$. Let us assume that the difference set of the sets is given by $\mathcal{A}_{{\bar{s}}_t^c(\ell)}\backslash \mathcal{A}_{{\bar{s}}_t(\ell)}=\big\{\delta_1(\ell),\dots,\delta_k(\ell)\big\}$. Then, according to equations \eqref{08010}, \eqref{08015}, \eqref{08031}, and \eqref{08033}, a simulation method that generates $\big(\pi_{\delta_k(\ell)}(\ell),\Sigma_{\delta_k(\ell)}(\ell)\big)$ as follows: if $r_{\delta(\ell)}>0$, then for $k=1,\dots,r_{\delta(\ell)}$, 
\begin{itemize}
\item generate $\Sigma_{\delta_k(\ell)}(\ell)$ from $\mathcal{IW}\Big(\nu_{0,\delta_k(\ell)},V_{0,\delta_k(\ell)}\Big)$,
\item generate $\pi_{\delta_k(\ell)}(\ell)$ from $\mathcal{N}\Big(\pi_{0,\delta_k(\ell)},\Lambda_{0,\delta_k(\ell)}\otimes\Sigma_{\delta_k(\ell)}(\ell)\Big)$.
\end{itemize}
Note that if $r_{\delta(\ell)}=0$, we do not need to generate the coefficient vector $\pi_{\delta(\ell)}(\ell)$ and covariance matrix $\Sigma_{\delta(\ell)}(\ell)$ from the density function $f\big(\pi_{\delta(\ell)},\Sigma_{\delta(\ell)}\big|\delta(\ell),\mathcal{F}_0\big)$. Let 
\begin{equation}\label{08181}
\hat{\beta}(\ell):=\big(\alpha_1(\ell),\dots,\alpha_{r_{\alpha}(\ell)}(\ell),\delta_1(\ell),\dots,\delta_{r_{\delta(\ell)}}(\ell)\big)'
\end{equation}
be an $(r_{\beta(\ell)}\times 1)$ regime vector. The regime vector $\hat{\beta}(\ell)$ has same elements as the duplication removed regime vector $\beta(\ell)$, but positions of the elements are different for the two regime vectors. Collect the realizations for $k=1,\dots,r_{\alpha(\ell)}$, $\Sigma_{\alpha_k(\ell)}(\ell)$ and $\pi_{\alpha_k(\ell)}(\ell)$ and for $k=1,\dots,r_{\delta(\ell)}$, $\Sigma_{\delta_k(\ell)}(\ell)$ and $\pi_{\delta_k(\ell)}(\ell)$ into $([ndr_{\beta(\ell)}]\times 1)$ vector $\pi_{\hat{\beta}(\ell)}(\ell)$ and $([nr_{\beta(\ell)}]\times n)$ matrix $\Sigma_{\hat{\beta}(\ell)}(\ell)$, namely,
\begin{eqnarray}\label{08112}
\pi_{\hat{\beta}(\ell)}(\ell)&:=&\Big(\pi_{\alpha_1(\ell)}(\ell)',\dots,\pi_{\alpha_{r_{\alpha(\ell)}}(\ell)}(\ell)',\pi_{\delta_1(\ell)}(\ell)',\dots,\pi_{\delta_{r_{\delta(\ell)}}(\ell)}(\ell)'\Big)'
\end{eqnarray}
and
\begin{eqnarray}\label{08113}
\Sigma_{\hat{\beta}(\ell)}(\ell)&:=&\Big[\Sigma_{\alpha_1(\ell)}(\ell):\dots:\Sigma_{\alpha_{r_{\alpha(\ell)}}(\ell)}(\ell),\Sigma_{\delta_1(\ell)}(\ell):\dots:\Sigma_{\delta_{r_{\delta(\ell)}}(\ell)}(\ell)\Big]'.
\end{eqnarray}
Similar to equation \eqref{08036}, for $u=t+1,\dots,T$, we denote a position of the regime $s_u(\ell)$ in the regime vector $\hat{\beta}(\ell)$ by $o_\ell(\ell)$. Let us define a matrix $D_{\hat{\beta}(\ell)}:=\big[j_{o_{t+1}(\ell)}:\dots:j_{o_T(\ell)}\big]'$, where $j_o(\ell)$ is an ($r_{\beta(\ell)}\times 1$) unit vector, whose $o$--th element 1 and others 0. Then, one revives the vector $\pi_{\bar{s}_t^c(\ell)}=\big(D_{\hat{\beta}(\ell)}\otimes I_{nd}\big)\pi_{\hat{\beta}(\ell)}$ and matrix
\begin{equation}\label{08114}
\Sigma_{\bar{s}_t^c(\ell)}=\text{diag}\big\{\big((D_{\hat{\beta}(\ell)}\otimes I_n)\Sigma_{\hat{\beta}(\ell)}\big)_1,\dots,\big((D_{\hat{\beta}(\ell)}\otimes I_n)\Sigma_{\hat{\beta}(\ell)}\big)_T\big\}.
\end{equation}

Fourth, for $\ell=1,\dots,\mathcal{L}$, we consider a simulation method that generates the regime vector $\bar{y}_t^c(\ell)$ from the density function $f(\bar{y}_t^c|\pi_{\bar{s}_t^c(\ell)}(\ell),\Sigma_{\bar{s}_t^c(\ell)}(\ell),\bar{s}_t^c(\ell),\mathcal{F}_t)$. Let us assume that
\begin{equation}\label{08115}
\varphi_s=\begin{bmatrix}
\varphi_{1,\bar{s}_t}\\ \varphi_{2,\bar{s}_t^c}
\end{bmatrix} ~~~\text{and}~~~
\Psi_s=\begin{bmatrix}
\Psi_{11,\bar{s}_t} & 0\\
\Psi_{21,\bar{s}_t^c} & \Psi_{22,\bar{s}_t^c}
\end{bmatrix}
\end{equation}
are partitions of the vector $\varphi_s$ and matrix $\Psi_s$, corresponding to random sub vectors $\bar{y}_t$ and $\bar{y}_t^c$ of the random vector $y$. Then, due to \citeA{Battulga24a}, a distribution of the random vector $\bar{y}_t^c$ is given by
\begin{eqnarray}\label{08116}
\bar{y}_t^c~|~\pi_{\beta},\Sigma_{\beta},s,\mathcal{F}_t &\sim& \mathcal{N}\Big(\Psi_{22,\bar{s}_t^c}^{-1}\big(\varphi_{2,\bar{s}_t^c}-\Psi_{21,\bar{s}_t^c}\bar{y}_t\big),\Psi_{22,\bar{s}_t^c}^{-1}\Sigma_{\bar{s}_t^c}(\Psi_{22,\bar{s}_t^c}^{-1})'\Big),
\end{eqnarray}
where $\Sigma_{\bar{s}_t^c}=\text{diag}\{\Sigma_{s_{t+1}},\dots,\Sigma_{s_T}\}$ is an $([n(T-t)]\times[n(T-t)])$ block diagonal matrix, corresponding to the regime vector $\bar{s}_t^c$. Thus, a simulation method that generate the vector of endogenous variables $\bar{y}_t^c(\ell)$ for given $\pi_{\bar{s}_t(\ell)}(\ell)$, $\Sigma_{\bar{s}_t(\ell)}(\ell)$, $s(\ell)$, and $\mathcal{F}_t$ as follows:
\begin{itemize}
\item generate $\bar{y}_t^c(\ell)$ from $\mathcal{N}\Big(\Psi_{22,\bar{s}_t^c(\ell)}^{-1}\big(\varphi_{2,\bar{s}_t^c(\ell)}-\Psi_{21,\bar{s}_t^c(\ell)}\bar{y}_t\big),\Psi_{22,\bar{s}_t^c(\ell)}^{-1}\Sigma_{\bar{s}_t^c(\ell)}(\ell)\big(\Psi_{22,\bar{s}_t^c(\ell)}^{-1}\big)'\Big),$
\end{itemize}
where the matrix $\Psi_{22,\bar{s}_t^c(\ell)}$ and vector $\varphi_{\bar{s}_t^c(\ell)}-\Psi_{21,\bar{s}_t^c(\ell)}\bar{y}_t$ are given by
\begin{equation}\label{08117}
\Psi_{22,\bar{s}_t^c(\ell)}:=\begin{bmatrix}
I_n & 0 & \dots & 0 & \dots & 0 & 0\\
-A_{1,s_{t+2}(\ell)} & I_n & \dots & 0 & \dots & 0 & 0\\
\vdots & \vdots & \dots & \vdots & \dots & \vdots & \vdots\\
0 & 0 & \dots & -A_{p-1,s_{T-1}(\ell)} & \dots & I_n & 0\\
0 & 0 & \dots & -A_{p,s_T(\ell)} & \dots & -A_{1,s_T(\ell)} & I_n
\end{bmatrix}
\end{equation}
and
\begin{equation}\label{08118}
\varphi_{2,\bar{s}_t^c(\ell)}-\Psi_{21,\bar{s}_t^c(\ell)}\bar{y}_t:=\begin{bmatrix}
A_{0,s_{t+1}(\ell)}\psi_{t+1}+A_{1,s_{t+1}(\ell)}y_t+\dots+A_{p,s_{t+1}(\ell)}y_{t+1-p}\\ 
A_{0,s_{t+2}(\ell)}\psi_{t+2}+A_{2,s_{t+2}(\ell)}y_t+\dots+A_{p,s_{t+2}(\ell)}y_{t+2-p}\\
\vdots\\
A_{0,s_{T-1}(\ell)}\psi_{T-1}\\
A_{0,s_T(\ell)}\psi_T
\end{bmatrix}
\end{equation}
and they are obtained from the vector $\pi_{\bar{s}_t^c(\ell)}(\ell)$. It should be noted that traditional methods that generate the vector $\bar{y}_t^c$ are based on an iterative method for $y_{t+1},\dots,y_T$ by generating $\xi_{t+1},\dots,\xi_T$, see \citeA{Karlsson13}. As a result, if $(T-t)$ is large, the simulation method reduces the computational burden that generates the random vector $\bar{y}_t^c$ as compared to the traditional algorithms.

\subsubsection{Importance Sampling Method}

Now, we consider the importance sampling method for the Bayesian MS--VAR process. We estimate a probability of a rare event, corresponding to the endogenous variables by the important sampling method. In the importance sampling method, one changes the real probability measure $\mathbb{P}$. The new probability measure $\tilde{\mathbb{P}}$ must be chosen that the rare event more frequently comes from than the real probability measure $\mathbb{P}$. Let $f(y,\pi_{\hat{s}},\Sigma_{\hat{s}},s,\mathsf{P}|\mathcal{F}_0)$ and $\tilde{f}(y,\pi_{\hat{s}},\Sigma_{\hat{s}},s,\mathsf{P}|\mathcal{F}_0)$ be joint density functions under the real probability measure $\mathbb{P}$ and new probability measure $\tilde{\mathbb{P}}$, respectively, for given initial information $\mathcal{F}_0$ and 
\begin{equation}\label{08119}
L=L(y,\pi_{\hat{s}},\Sigma_{\hat{s}},s,\mathsf{P}|\mathcal{F}_0)=f(y,\pi_{\hat{s}},\Sigma_{\hat{s}},s,\mathsf{P}|\mathcal{F}_0)/\tilde{f}(y,\pi_{\hat{s}},\Sigma_{\hat{s}},s,\mathsf{P}|\mathcal{F}_0)
\end{equation}
be the likelihood ratio. Let us choose density functions that corresponds to the new probability measure $\mathbb{\tilde{P}}$ by 
\begin{equation}\label{08120}
\tilde{f}(\bar{y}_t,\Sigma_{\hat{s}},s,\mathsf{P}|\mathcal{F}_0)=f(\bar{y}_t,\Sigma_{\hat{s}},s,\mathsf{P}|\mathcal{F}_0),
\end{equation}
if $s_u\in\mathcal{A}_{\bar{s}_t}\cap\mathcal{A}_{\bar{s}_t^c}$, then
\begin{eqnarray}\label{08121}
&&\tilde{f}(\pi_{\hat{s}}|\Sigma_{\hat{s}},s,\mathsf{P},\mathcal{F}_t)=f(\pi_{\hat{s}}|\Sigma_{\hat{s}},s,\mathsf{P},\mathcal{F}_t)=f(\pi_{\hat{s}}|\Sigma_{\hat{s}},s,\mathcal{F}_t)\nonumber\\
&&\Big/\frac{1}{(2\pi)^{nd}|A_{s_u|u}|^{1/2}}\exp\bigg\{-\frac{1}{2}\Big(\pi_{s_u}-\pi_{0,s_u|u}\Big)'A_{s_u|u}^{-1}\Big(\pi_{s_u}-\pi_{0,s_u|u}\Big)\bigg\}\nonumber\\
&&\times\frac{1}{(2\pi)^{nd}|A_{s_u|u}|^{1/2}}\exp\bigg\{-\frac{1}{2}\Big(\pi_{s_u}-\pi_{0,s_u|u}-\theta_{s_u} A_{s_u|u}(\mathsf{Y}_u\otimes I_n)z_u\Big)'\\
&&\times A_{s_u|u}^{-1}\Big(\pi_{s_u}-\pi_{0,s_u|u}-\theta_{s_u} A_{s_u|u}(\mathsf{Y}_u\otimes I_n)z_u\Big)\bigg\}\nonumber
\end{eqnarray}
if $s_u\in\mathcal{A}_{\bar{s}_t^c}\backslash\mathcal{A}_{\bar{s}_t}$, then
\begin{eqnarray}\label{08122}
&&\tilde{f}(\pi_{\hat{s}}|\Sigma_{\hat{s}},s,\mathsf{P},\mathcal{F}_t)=f(\pi_{\hat{s}}|\Sigma_{\hat{s}},s,\mathsf{P},\mathcal{F}_t)=f(\pi_{\hat{s}}|\Sigma_{\hat{s}},s,\mathcal{F}_t)\nonumber\\
&&\Big/\frac{1}{(2\pi)^{nd}|A_{s_u}|^{1/2}}\exp\bigg\{-\frac{1}{2}\Big(\pi_{s_u}-\pi_{0,s_u}\Big)'A_{s_u}^{-1}\Big(\pi_{s_u}-\pi_{0,s_u}\Big)\bigg\}\nonumber\\
&&\times\frac{1}{(2\pi)^{nd}|A_{s_u}|^{1/2}}\exp\bigg\{-\frac{1}{2}\Big(\pi_{s_u}-\pi_{0,s_u}-\theta_{s_u} A_{s_u}(\mathsf{Y}_u\otimes I_n)z_u\Big)'\\
&&\times A_{s_u}^{-1}\Big(\pi_{s_u}-\pi_{0,s_u}-\theta_{s_u} A_{s_u}(\mathsf{Y}_u\otimes I_n)z_u\Big)\bigg\}\nonumber
\end{eqnarray}
and
\begin{eqnarray}\label{08123}
&&\tilde{f}(\bar{y}_t^c|\pi_{\hat{s}},\Sigma_{\hat{s}},s,\mathsf{P},\mathcal{F}_t)=f(\bar{y}_t^c|\pi_{\hat{s}},\Sigma_{\hat{s}},s,\mathsf{P},\mathcal{F}_t)=f(\bar{y}_t^c|\pi_{\hat{s}},\Sigma_{\hat{s}},s,\mathcal{F}_t)\nonumber\\
&&\Big/\frac{1}{(2\pi)^n|\Sigma_{s_u}|^{1/2}}\exp\bigg\{-\frac{1}{2}\Big(y_u-\Pi_{s_u}\mathsf{Y}_u\Big)'\Sigma_{s_u}^{-1}\Big(y_u-\Pi_{s_u}\mathsf{Y}_u\Big)\bigg\}\\
&&\times\frac{1}{(2\pi)^n|\Sigma_{s_u}|^{1/2}}\exp\bigg\{-\frac{1}{2}\Big(y_u-\Pi_{s_u}\mathsf{Y}_u-\theta_{s_u}\Sigma_{s_u}z_u\Big)'\Sigma_{s_u}^{-1}\Big(y_u-\Pi_{s_u}\mathsf{Y}_u-\theta_{s_u}\Sigma_{s_u}z_u\Big)\bigg\}\nonumber
\end{eqnarray}
for $u=t+1,\dots,T$, where $A_{s_u}:=(\Lambda_{0,s_u}\otimes \Sigma_{s_u})$ is an $([nd]\times [nd])$ matrix, $\theta_{s_u}$ is a positive constant, depending on the random covariance matrix $\Sigma_{s_t}$, and regime $s_u$ and $z_u$ is an $(n\times 1)$ vector, whose elements are known. Note that if $\theta_{s_u}=0$, then the new probability measure $\tilde{\mathbb{P}}$ equals the real probability measure $\mathbb{P}$. If we compare the density functions $\tilde{f}(\bar{y}_t^c|\pi_{\hat{s}},\Sigma_{\hat{s}},s,\mathsf{P},\mathcal{F}_t)$ and $f(\bar{y}_t^c|\pi_{\hat{s}},\Sigma_{\hat{s}},s,\mathsf{P},\mathcal{F}_t)$, one can conclude that conditional distribution of $y_u$ changes from $y_u|\pi_{s_u},s_u,\mathcal{F}_{u-1}\sim\mathcal{N}(\Pi_{s_u}\mathsf{Y}_u,\Sigma_{s_u})$ to $y_u|\pi_{s_u},s_u,\mathcal{F}_{u-1}\sim\mathcal{N}(\Pi_{s_u}\mathsf{Y}_u+\theta_{s_u}\Sigma_{s_u}z_u,\Sigma_{s_u})$ and for each $v=t+1,\dots,T$ $(v\neq u)$, the conditional distribution of other sub random vector $y_v$  of the random vector $\bar{y}_t^c$ does not change. The same explanation holds for the density functions $\tilde{f}(\pi_{\hat{s}}|\Sigma_{\hat{s}},s,\mathsf{P},\mathcal{F}_t)$ and $f(\pi_{\hat{s}}|\Sigma_{\hat{s}},s,\mathsf{P},\mathcal{F}_t)$. Then, it can be shown that for $u=t+1,\dots,T$, the likelihood ratio is given by
\begin{equation}\label{08124}
L_u=\exp\big\{-\theta_{s_u} X_u+\psi(\theta_{s_u})\big\}.
\end{equation}
where the random variable $X_u$ is given by $X_u:=z_u'y_u$ and the quadratic function $\psi(\theta_{s_u})$ for $\theta_{s_u}$ is given by 
\begin{equation}\label{08125}
\psi(\theta_{s_u}):=\begin{cases}
\theta_u z_u'\pi_{0,s_u|u}^\circ\mathsf{Y}_u+\frac{1}{2}\theta_u^2\big(1+\mathsf{Y}_u'\Lambda_{0,s_u|u}\mathsf{Y}_u\big)z_u'\Sigma_{s_u} z_u & \text{if}~~~s_u\in\mathcal{A}_{\bar{s}_t}\cap\mathcal{A}_{\bar{s}_t^c},\\
\theta_u z_u'\pi_{0,s_u}^\circ\mathsf{Y}_u+\frac{1}{2}\theta_u^2\big(1+\mathsf{Y}_u'\Lambda_{0,s_u}\mathsf{Y}_u\big)z_u'\Sigma_{s_u} z_u & \text{if}~~~s_u\in \mathcal{A}_{\bar{s}_t^c}\backslash\mathcal{A}_{\bar{s}_t}.
\end{cases}
\end{equation}
For each $u=t+1,\dots,T$, by choosing $z_u$ by the unit vector, one can extract elements of the vector of endogenous variables $y_u$. If the process $y_t$ consists of returns of financial assets, then by choosing $z_u$ by weight vector, one obtains portfolio return. 

Now we consider a conditional probability $\mathbb{P}(X_u>x_u|\pi_\beta,\Sigma_\beta,s,\mathcal{F}_{u-1})$ for large $x_u\in\mathbb{R}$ and $u=t+1,\dots,T$. Since $\theta_{s_u}$ is the positive constant, for the conditional probability, by equation \eqref{08124}, the following inequality holds
\begin{equation}\label{08126}
\mathbb{P}(X_u>x_u|\pi_\beta,\Sigma_\beta,s,\mathcal{F}_{u-1})=\mathbb{\tilde{E}}\big[1_{\{X_u>x_u\}}L_u\big|\pi_\beta,\Sigma_\beta,s,\mathcal{F}_{u-1}\big]\leq \exp\big\{-\theta_{s_u} x_u+\psi(\theta_{s_u})\big\}
\end{equation}
for $u=t+1,\dots,T$, where for a generic event $A\in \mathcal{H}_T$, $1_A$ is the indicator function of the event $A$, see \citeA{Glasserman00}. Also, for the second order moment, it holds
\begin{equation}\label{08127}
m_2(x_u,\theta_{s_u}):=\mathbb{\tilde{E}}\big[1_{\{X_u>x_u\}}L_u^2\big|\pi_\beta,\Sigma_\beta,s,\mathcal{F}_{u-1}\big]\leq \exp\big\{-2\theta_{s_u} x_u+2\psi(\theta_{s_u})\big\}.
\end{equation}
To reduce a variance of the importance sampling, we need to keep the right hand side of the above inequality as low as possible. To minimize the right--hand side of the above equation, we minimize the exponent using the parameter $\theta_{s_t}$. The minimizer of the right--hand side of the above equation is given by 
\begin{equation}\label{08128}
\theta_{s_u}(x_u):=\begin{cases}
\displaystyle\frac{x_u-z_u'\pi_{0,s_u|u}^\circ\mathsf{Y}_u}{\big(1+\mathsf{Y}_u'\Lambda_{0,s_u|u}\mathsf{Y}_u\big)z_u'\Sigma_{s_u}z_u} & \text{if}~~~s_u\in \mathcal{A}_{\bar{s}_t}\cap \mathcal{A}_{\bar{s}_t^c},\\
\displaystyle\frac{x_u-z_u'\pi_{0,s_u}^\circ\mathsf{Y}_u}{\big(1+\mathsf{Y}_u'\Lambda_{0,s_u}\mathsf{Y}_u\big)z_u'\Sigma_{s_u}z_u} & \text{if}~~~s_u\in \mathcal{A}_{\bar{s}_t^c}\backslash \mathcal{A}_{\bar{s}_t}
\end{cases}
\end{equation}
for $u=t+1,\dots,T$. Therefore, an importance sampling method that estimates the conditional probabilities $\mathbb{P}(X_u>x_u|\mathcal{F}_t)$ for $u=t+1,\dots,T$ as follows: for $\ell=1,\dots,\mathcal{L}$,
\begin{itemize}
\item generate $\big(\bar{y}_t^c(\ell),\pi_{\bar{s}_t^c(\ell)}(\ell),\Sigma_{\bar{s}_t^c(\ell)}(\ell),\bar{s}_t^c(\ell)\big)$ using the general simulation method,
\item for $u=t+1,\dots,T$,
\begin{itemize}
\item calculate $\theta_{s_u(\ell)}^*(x_u,\ell)$ using equation \eqref{08128},
\item if $s_u(\ell)\in \mathcal{A}_{\bar{s}_t(\ell)}\cap \mathcal{A}_{\bar{s}_t^c(\ell)}$, generate $\pi_{s_u(\ell)}^*(\ell)$ from 
\begin{equation}\label{08129}
\mathcal{N}\Big(\pi_{0,s_u(\ell)|u}(\ell)+\theta_{s_u(\ell)}^*(x_u,\ell)A_{s_u(\ell)|u}(\ell)(\mathsf{Y}_u(\ell)\otimes I_n)z_u,A_{s_u(\ell)|u}(\ell)\Big),
\end{equation}
\item if $s_u(\ell)\in \mathcal{A}_{\bar{s}_t^c(\ell)}\backslash \mathcal{A}_{\bar{s}_t(\ell)}$, generate $\pi_{s_u(\ell)}^*(\ell)$ from 
\begin{equation}\label{08130}
\mathcal{N}\Big(\pi_{0,s_u(\ell)}(\ell)+\theta_{s_u(\ell)}^*(x_u,\ell)A_{s_u(\ell)}(\ell)(\mathsf{Y}_u(\ell)\otimes I_n)z_u,A_{s_u(\ell)}(\ell)\Big),
\end{equation}
\item generate $y_u^*(\ell)$ from 
\begin{equation}\label{08131}
\mathcal{N}\Big(\Pi_{s_u}^*\mathsf{Y}_u(\ell)+\theta_{s_u(\ell)}^*(x_u,\ell)\Sigma_{s_u(\ell)}z_u,\Sigma_{s_u(\ell)}(\ell)\Big), 
\end{equation}
where $\pi_{s_u(\ell)}^*(\ell)=\text{vec}(\Pi_{s_u(\ell)}^*(\ell))$,
\item calculate $L_u^*(\ell)$ using equation \eqref{08124}, where $X_u^*(\ell)=z_u'y_u^*(\ell)$,
\end{itemize}
\item and for $u=t+1,\dots,T$, estimate the probabilities $\mathbb{P}(X_u>x_u|\mathcal{F}_t)$ by
\begin{equation}\label{08132}
\hat{\mathbb{P}}(X_u>x_u|\mathcal{F}_t)=\frac{1}{\mathcal{L}}\sum_{\ell=1}^{\mathcal{L}}1_{\{X_u^*(\ell)>x_u\}}L_u^*(\ell).
\end{equation}
\end{itemize}

\section{Conclusion}

In this paper, for the general Bayesian MS--VAR process, we obtain some useful density functions for Monte--Carlo simulations. The density functions tell us that conditional on the regime vectors and initial information, the vector of endogenous variables is independent of model's some random components. Thus, one only needs the prior distributions to calculate the density functions, and the results have yet to be explored before.

In a particular case of the general Bayesian MS--VAR process, we also get closed--form density functions of the random components of the model. In particular, we find that joint distribution of future values of the random coefficient matrix is a product of matrix variate student distributions, see equation \eqref{08047}. Hence, one can analyze impulse response by directly generating the coefficient matrices from the distribution functions. Also, we provide a new density function, which has yet to be introduced before of future values of the endogenous variables; see equations \eqref{08043} and \eqref{08071}. Thus, future studies may concentrate on marginal density functions and direct simulation methods for the density function. Further, we obtain a characteristic function of the random coefficient matrix, which can be used to calculate the forecast of the endogenous variables.

In the paper, we develop Monte--Carlo simulation algorithms. The simulation method's novelty is that it removes duplication in a regime vector. As a result, our proposed Monte--Carlo simulation method departs from the previous simulation methods with regime switching. We also provide importance sampling method to estimate probability of a rare event, corresponding to the future endogenous variables. Since the method can be used to calculate quantiles, in this case, the quantiles of the future endogenous variables become more reliable than navy simulation methods. To obtain smoothed probability inference, \citeA{Kim94}'s smoothing method is widely used in practice. But this method is based on approximation. For this reason, we introduce a very simple, exact (in some special cases), and new smoothing method.

\section*{Appendix A: Proofs of Propositions}

\begin{proof}[\textbf{Proof of Proposition \ref{prop01}}]
Let us consider a joint density function of the random vectors $\bar{y}_t$ and $s$ and random matrices $\Pi_{\hat{s}}$, $\Gamma_{\hat{s}}$, and $\mathsf{P}$ for given initial information $\mathcal{F}_0$. According to the conditional probability formula, one gets that
\begin{equation}\label{08020}
f(\bar{y}_t,\Pi_{\hat{s}},\Gamma_{\hat{s}},s,\mathsf{P}|\mathcal{F}_0)=f(\bar{y}_t|\Pi_{\hat{s}},\Gamma_{\hat{s}},s,\mathsf{P},\mathcal{F}_0)f(\Pi_{\hat{s}},\Gamma_{\hat{s}}|s,\mathsf{P},\mathcal{F}_0)f(s|\mathsf{P},\mathcal{F}_0)f(\mathsf{P}|\mathcal{F}_0).
\end{equation}
Since the random vector $\bar{y}_t$ depends on $\Pi_{\alpha}$, $\Gamma_{\alpha}$, $\bar{s}_t$, the first joint density function of the right--hand side of the above equation equals $f(\bar{y}_t|\Pi_{\alpha},\Gamma_{\alpha},\bar{s}_t,\mathcal{F}_0)$. As the random coefficient matrix $(\Pi_{\hat{s}},\Gamma_{\hat{s}})$ is independent of the transition probability matrix $\mathsf{P}$ conditional on $s$ and $\mathcal{F}_0$, the second joint density function of the right--hand side can be represented by equation \eqref{08011}. According to the Markov property \eqref{08004}, the third joint density function equals $f(\bar{s}_t|\mathsf{P},\mathcal{F}_0)f(\bar{s}_t^c|\bar{s}_t,\mathsf{P},\mathcal{F}_0)$. Consequently, one obtains equation \eqref{08013}. By the conditional probability formula, we have that
\begin{eqnarray}\label{08166}
f(\bar{s}_t^c|\Pi_\alpha,\Gamma_\alpha,\bar{s}_t,\mathsf{P},\mathcal{F}_t)=\frac{f(\bar{y}_t|\Pi_\alpha,\Gamma_\alpha,s,\mathsf{P},\mathcal{F}_0)f(\mathsf{P}|\Pi_\alpha,\Gamma_\alpha,s,\mathcal{F}_0)f(\Pi_\alpha,\Gamma_\alpha|s,\mathcal{F}_0)f(s|\mathcal{F}_0)}{f(\bar{y}_t|\Pi_\alpha,\Gamma_\alpha,\bar{s}_t,\mathsf{P},\mathcal{F}_0)f(\mathsf{P}|\Pi_\alpha,\Gamma_\alpha,\bar{s}_t,\mathcal{F}_0)f(\Pi_\alpha,\Gamma_\alpha|\bar{s}_t,\mathcal{F}_0)f(\bar{s}_t|\mathcal{F}_0)}
\end{eqnarray}
The first density functions in the nominator and denominator are equal to $f(\bar{y}_t|\Pi_\alpha,\Gamma_\alpha,\bar{s}_t,\mathcal{F}_0)$. The third density functions in the nominator equal denominator are equal to each other. Consequently, since the transition probability matrix $\mathsf{P}$ is independent of coefficient matrix $(\Pi_\alpha,\Sigma_\alpha)$ for given regime vector $s$ and initial information $\mathcal{F}_0$, one obtains equation \eqref{08018}.  If we integrate equation \eqref{08013} by $\mathsf{P}$, then one finds that
\begin{equation}\label{08021}
f(\bar{y}_t,\Pi_{\hat{s}},\Gamma_{\hat{s}},s|\mathcal{F}_0)=f\big(\bar{y}_t,\Pi_{\alpha},\Gamma_{\alpha},\bar{s}_t\big|\mathcal{F}_0\big)f_*\big(\Pi_{\delta},\Gamma_{\delta}\big|\delta,\mathcal{F}_0\big)f(\bar{s}_t^c|\bar{s}_t,\mathcal{F}_0).
\end{equation}
By integrating the above equation with respect to $(\Pi_{\hat{s}},\Gamma_{\hat{s}})$, we get that
\begin{equation}\label{08023}
f(\bar{y}_t,s|\mathcal{F}_0)=f(\bar{y}_t,\bar{s}_t|\mathcal{F}_0)f(\bar{s}_t^c|\bar{s}_t,\mathcal{F}_0).
\end{equation}
If we divide equation \eqref{08021} by the above equation, then one obtains 
\begin{equation}\label{08024}
f\big(\Pi_{\hat{s}},\Gamma_{\hat{s}}\big|s,\mathcal{F}_t\big)=f\big(\Pi_{\alpha},\Gamma_{\alpha}\big|\bar{s}_t,\mathcal{F}_t\big)f_*\big(\Pi_{\delta},\Gamma_{\delta}\big|\delta,\mathcal{F}_0\big).
\end{equation}
Integrating the above equation by $(\Pi_\delta,\Gamma_\delta)$, we have $f\big(\Pi_{\alpha},\Gamma_{\alpha}\big|s,\mathcal{F}_t\big)=f\big(\Pi_{\alpha},\Gamma_{\alpha}\big|\bar{s}_t,\mathcal{F}_t\big)$.
Consequently, due to the conditional probability formula, we have that
\begin{equation}\label{08025}
f\big(\Pi_{\hat{s}},\Gamma_{\hat{s}}\big|s,\mathcal{F}_t\big)=f\big(\Pi_{\alpha},\Gamma_{\alpha}\big|\bar{s}_t,\mathcal{F}_t\big)f_*\big(\Pi_{\delta},\Gamma_{\delta}\big|\Pi_{\alpha},\Gamma_{\alpha},s,\mathcal{F}_0\big).
\end{equation}
Thus, equating equation \eqref{08024} with the above equation, we get that
\begin{equation}\label{08167}
f\big(\Pi_{\delta},\Gamma_{\delta}\big|\Pi_{\alpha},\Gamma_{\alpha},s,\mathcal{F}_t\big)=f_*\big(\Pi_{\delta},\Gamma_{\delta}\big|\delta,\mathcal{F}_0\big).
\end{equation}
By the conditional probability formula, we have that
\begin{eqnarray}\label{08168}
f\big(\Pi_{\delta},\Gamma_{\delta}\big|\Pi_{\alpha},\Gamma_{\alpha},s,\mathsf{P},\mathcal{F}_t\big)=\frac{f(\bar{y}_t|\Pi_{\hat{s}},\Gamma_{\hat{s}},s,\mathsf{P},\mathcal{F}_0)f(\Pi_{\hat{s}},\Gamma_{\hat{s}}|s,\mathsf{P},\mathcal{F}_0)f(s,\mathsf{P}|\mathcal{F}_0)}{f(\bar{y}_t|\Pi_{\alpha},\Gamma_{\alpha},s,\mathsf{P},\mathcal{F}_0)f(\Pi_{\alpha},\Gamma_{\alpha}|s,\mathsf{P},\mathcal{F}_0)f(s,\mathsf{P}|\mathcal{F}_0)}.
\end{eqnarray}
The first density functions in the nominator and denominator are equal to $f(\bar{y}_t|\Pi_\alpha,\Gamma_\alpha,\bar{s}_t,\mathcal{F}_0)$. The second density functions in the nominator equal denominator do not depend on the transition probability matrix $\mathsf{P}$. Thus, according to equation \eqref{08011}, we get equation \eqref{08015}. If we integrate equation \eqref{08024} by $(\Pi_\epsilon,\Gamma_\epsilon)$, then by equation \eqref{08007} and \eqref{08009}, we obtain equation \eqref{08016}. Due to the conditional probability formula, we have that
\begin{eqnarray}\label{08169}
f\big(\mathsf{P}|\Pi_{\alpha},\Gamma_{\alpha},\bar{s}_t,\mathcal{F}_t\big)=\frac{f(\bar{y}_t,\Pi_{\alpha},\Gamma_{\alpha}|\bar{s}_t,\mathsf{P},\mathcal{F}_0)f(\bar{s}_t,\mathsf{P}|\mathcal{F}_0)}{f(\bar{y}_t,\Pi_{\alpha},\Gamma_{\alpha}|\bar{s}_t,\mathcal{F}_0)f(\bar{s}_t|\mathcal{F}_0)}.
\end{eqnarray}
Because the first density function in nominator is $f(\bar{y}_t,\Pi_{\alpha},\Gamma_{\alpha}|\bar{s}_t,\mathcal{F}_0)$, we get equation \eqref{08017}. Finally, according to the conditional probability formula, we have that
\begin{eqnarray}\label{08170}
f\big(\Pi_{\alpha},\Gamma_{\alpha}|\bar{s}_t,\mathsf{P},\mathcal{F}_t\big)=\frac{f(\bar{y}_t,\Pi_{\alpha},\Gamma_{\alpha}|\bar{s}_t,\mathsf{P},\mathcal{F}_0)f(\mathsf{P}|\bar{s}_t,\mathcal{F}_0)f(\bar{s}_t|\mathcal{F}_0)}{f(\mathsf{P}|\bar{s}_t,\mathcal{F}_t)f(\bar{y}_t|\bar{s}_t,\mathcal{F}_0)f(\bar{s}_t|\mathcal{F}_0)}.
\end{eqnarray}
Since the first density function in nominator does not depend on the transition probability matrix $\mathsf{P}$ and the first density function in denominator does not depend on the random vector $\bar{y}_t$, one obtains equation \eqref{08019}. That completes the proof. 
\end{proof}

\begin{proof}[\textbf{Proof of Proposition \ref{prop02}}]
First, since $(\pi_{\alpha_1},\Sigma_{\alpha_1}),\dots,(\pi_{r_{\alpha}},\Sigma_{r_{\alpha}})$ are independent for given regime vector $\bar{s}_t$ and initial information $\mathcal{F}_0$, observe that conditional density functions of the random vectors $\bar{y}$ and $\pi_{\alpha}$ and random matrix $\Sigma_{\alpha}$ are given by
\begin{eqnarray}\label{08048}
f(\bar{y}_t|\pi_{\alpha},\Sigma_{\alpha},\bar{s}_t,\mathcal{F}_0)&=&\frac{1}{(2\pi)^{nt/2}}\prod_{k=1}^{r_{\alpha}}|\Sigma_{\alpha_k}|^{-q_{t,\alpha_k}/2}\nonumber\\
&\times&\exp\bigg\{-\frac{1}{2}\sum_{k=1}^{r_{\alpha}}\Big(y_{t,\alpha_k}-\big((\mathsf{Y}_{t,\alpha_k}^\circ)'\otimes I_{n}\big)\pi_{\alpha_k}\Big)'\\
&\times&\big(I_{q_{t,\alpha_k}}\otimes \Sigma_{\alpha_k}^{-1}\big)\Big(y_{t,\alpha_k}-\big((\mathsf{Y}_{t,\alpha_k}^\circ)'\otimes I_{n}\big)\pi_{\alpha_k}\Big)\bigg\},\nonumber
\end{eqnarray}
\begin{eqnarray}\label{08049}
f(\pi_{\alpha}|\Sigma_{\alpha},\bar{s}_t,\mathcal{F}_0)&=&\frac{1}{(2\pi)^{nr_{\alpha}d/2}\prod_{k=1}^{r_\alpha}|\Lambda_{0,\alpha_k}|^{n/2}}\prod_{k=1}^{r_{\alpha}}|\Sigma_{\alpha_k}|^{-d/2}\\
&\times&\exp\bigg\{-\frac{1}{2}\sum_{k=1}^{r_{\alpha}}\big(\pi_{\alpha_k}-\pi_{0,\alpha_k}\big)'\big(\Lambda_{0,\alpha_k}^{-1}\otimes \Sigma_{\alpha_k}^{-1}\big)\big(\pi_{\alpha_k}-\pi_{0,\alpha_k}\big)\bigg\},\nonumber
\end{eqnarray}
and
\begin{equation}\label{08050}
f(\Sigma_{\alpha}|\bar{s}_t,\mathcal{F}_0)=\prod_{k=1}^{r_{\alpha}}\frac{|V_{0,\alpha_k}|^{\nu_{0,\alpha_k}/2}}{\Gamma_{n}(\nu_{0,\alpha_k}/2)2^{n\nu_{0,\alpha_k}/2}}|\Sigma_{\alpha_k}|^{-(\nu_{0,\alpha_k}+n+1)/2}\exp\bigg\{-\frac{1}{2}\text{tr}\Big(V_{0,\alpha_k}\Sigma_{\alpha_k}^{-1}\Big)\bigg\},
\end{equation} 
respectively. Consequently, by the completing square method, a joint conditional density function of the random vectors $\bar{y}_t$ and $\pi_{\alpha}$ is
\begin{eqnarray}\label{08051}
&&f(\bar{y}_t,\pi_{\alpha}|\Sigma_{\alpha},\bar{s}_t,\mathcal{F}_0)\nonumber\\
&&= c_1\prod_{k=1}^{r_{\alpha}}|\Sigma_{\alpha_k}|^{-(q_{t,\alpha_k}+d)/2}\exp\bigg\{-\frac{1}{2}\sum_{k=1}^{r_{\alpha}}\Big(\pi_{\alpha_k}-\pi_{0,\alpha_k|t}\Big)'A_{\alpha_k|t}^{-1}\Big(\pi_{\alpha_k}-\pi_{0,\alpha_k|t}\Big)\bigg\}\\
&&\times\exp\bigg\{-\frac{1}{2}\sum_{k=1}^{r_{\alpha}}\Big(y_{t,\alpha_k}'\big(I_{q_{t,\alpha_k}}\otimes \Sigma_{\alpha_k}^{-1}\big)y_{t,\alpha_k}+\pi_{0,\alpha_k}'\big(\Lambda_{0,\alpha_k}^{-1}\otimes \Sigma_{\alpha_k}^{-1}\big)\pi_{0,\alpha_k}-\pi_{0,\alpha_k|t}'A_{\alpha_k|t}^{-1}\pi_{0,\alpha_k|t}\Big)\bigg\},\nonumber
\end{eqnarray}
where normalizing coefficient equals
\begin{equation}\label{08052}
c_1:=\frac{1}{(2\pi)^{n(t+r_{\alpha}d)/2}\prod_{k=1}^{r_{\alpha}}|\Lambda_{0,\alpha_k}|^{n/2}}.
\end{equation}
If we integrate the above joint density function with respect to the vector $\pi_{\alpha}$, then an integral, corresponding to the first exponential is proportional to $\prod_{k=1}^{r_{\alpha}}|A_{\alpha_k|t}|^{1/2}=\prod_{k=1}^{r_{\alpha}}|(\mathsf{Y}_{t,\alpha_k}^\circ(\mathsf{Y}_{t,\alpha_k}^\circ)'+\Lambda_{0,\alpha_k}^{-1})|^{-n/2} |\Sigma_{\alpha_k}|^{d/2}$. Therefore, we have that
\begin{eqnarray}\label{08053}
&&f(\bar{y}_t|\Sigma_{\alpha},\bar{s}_t,\mathcal{F}_0)= c_2\prod_{k=1}^{r_{\alpha}}|\Sigma_{\alpha_k}|^{-q_{t,\alpha_k}/2}\\
&&\times\exp\bigg\{-\frac{1}{2}\sum_{k=1}^{r_{\alpha}}\Big(y_{t,\alpha_k}'\big(I_{q_{t,\alpha_k}}\otimes \Sigma_{\alpha_k}^{-1}\big)y_{t,\alpha_k}+\pi_{0,\alpha_k}'\big(\Lambda_{0,\alpha_k}^{-1}\otimes \Sigma_{\alpha_k}^{-1}\big)\pi_{0,\alpha_k}-\pi_{0,\alpha_k|t}'A_{\alpha_k|t}^{-1}\pi_{0,\alpha_k|t}\Big)\bigg\},\nonumber
\end{eqnarray}
where the normalizing coefficient equals
\begin{equation}\label{08054}
c_2:=\frac{1}{(2\pi)^{nt/2}\prod_{k=1}^{r_{\alpha}}|\Lambda_{0,\alpha_k}|^{n/2}|\Lambda_{0,\alpha_k|t}^{-1}|^{n/2}}.
\end{equation}
Hence, according to the well--known formula that for suitable matrices $A,B,C,D$, 
\begin{equation}\label{08055}
\text{vec}(A)'(B\otimes C)\text{vec}(D)=\text{tr}(DB'A'C),
\end{equation}
we find that
\begin{equation}\label{08056}
f(\bar{y}_t|\Sigma_{\alpha},\bar{s}_t,\mathcal{F}_0)=c_2\prod_{k=1}^{r_{\alpha}}|\Sigma_{\alpha_k}|^{-q_{t,\alpha_k}/2}\exp\bigg\{-\frac{1}{2}\sum_{k=1}^{r_{\alpha}}\text{tr}\big(B_{t,\alpha_k}\Sigma_{\alpha_k}^{-1}\big)\bigg\}.
\end{equation}
Thus, it follows from equations \eqref{08050} and \eqref{08056} that a joint conditional density function of the random vector $\bar{y}_t$ and random matrix $\Sigma_{\alpha}$ is  
\begin{equation}\label{08057}
f(\bar{y}_t,\Sigma_{\alpha}|\bar{s}_t,\mathcal{F}_0)= c_3\prod_{k=1}^{r_{\alpha}}|\Sigma_{\alpha_k}|^{-(\nu_{0,\alpha_k|t}+n+1)/2}\exp\bigg\{-\frac{1}{2}\sum_{k=1}^{r_{\alpha}}\text{tr}\Big(\big(B_{t,\alpha_k}+V_{0,\alpha_k}\big)\Sigma_{\alpha_k}^{-1}\Big)\bigg\},
\end{equation}
where the normalizing coefficient equals
\begin{equation}\label{08058}
c_3:=\frac{1}{(2\pi)^{nt/2}}\prod_{k=1}^{r_{\alpha}}\frac{|V_{0,\alpha_k}|^{\nu_{0,\alpha_k}/2}}{|\Lambda_{0,\alpha_k}|^{n/2}|\Lambda_{0,\alpha_k|t}^{-1}|^{n/2}\Gamma_{n}(\nu_{0,\alpha_k}/2)2^{n\nu_{0,\alpha_k}/2}}.
\end{equation}
Consequently, a conditional density function of the random vector $\bar{y}_t$ is given by
\begin{eqnarray}\label{08059}
f(\bar{y}_t|\bar{s}_t,\mathcal{F}_0)&=&\int_{\Sigma_{\alpha_1},\dots,\Sigma_{\alpha_{r_{\alpha}}}>0}f(\bar{y}_t,\Sigma_{\alpha}|\bar{s}_t,\mathcal{F}_0)d\Sigma_{\alpha_1}\dots d\Sigma_{\alpha_{r_{\alpha}}}\nonumber\\
&=&c_3\prod_{k=1}^{r_{\alpha}}\frac{\Gamma_{n}\big(\nu_{0,\alpha_k|t}/2\big)2^{n\nu_{0,\alpha_k|t}/2}}{\big|B_{t,\alpha_k}+V_{0,\alpha_k}\big|^{\nu_{0,\alpha_k|t}/2}}.
\end{eqnarray}
If we divide equations \eqref{08051} and \eqref{08057} by equations \eqref{08053} and \eqref{08059}, respectively, then one obtains equations \eqref{08045} and \eqref{08046}. According to equation \eqref{08016}, we have that
\begin{equation}\label{08060}
f(\pi_\beta,\Sigma_\beta|s,\mathcal{F}_t)=f(\pi_\gamma,\Sigma_\gamma|\bar{s}_t,\mathcal{F}_t)f_*(\pi_\delta,\Sigma_\delta|\delta,\mathcal{F}_0).
\end{equation}
We consider the first joint conditional density of the right--hand side of the above equation. By integrating a product of density functions \eqref{08045} and \eqref{08046} by $(\pi_\epsilon,\Sigma_\epsilon)$ and taking account that
\begin{equation}\label{08061}
\sum_{k=1}^{r_{\gamma}}\Big(\pi_{\gamma_k}-\pi_{0,\gamma_k|t}^\circ\Big)'A_{\gamma_k|t}^{-1}\Big(\pi_{\gamma_k}-\pi_{0,\gamma_k|t}^\circ\Big)=\sum_{k=1}^{r_\gamma}\mathrm{tr}\Big((\pi_{\gamma_k}^\circ-\pi_{0,\gamma_k|t}^\circ)\Lambda_{0,\gamma_k|t}^{-1}(\pi_{\gamma_k}^\circ-\pi_{0,\gamma_k|t}^\circ)'\Sigma_{\gamma_k}^{-1}\Big)
\end{equation}
the joint conditional density function is
\begin{eqnarray}\label{08062}
&&f(\pi_\gamma,\Sigma_{\gamma}|\bar{s}_t,\mathcal{F}_t)=c_4
\prod_{k=1}^{r_{\gamma}}|\Sigma_{\gamma_k}|^{-(\nu_{0,\gamma_k|t}+d+n+1)/2}\\
&&\times\exp\bigg\{-\frac{1}{2}\sum_{k=1}^{r_{\gamma}}\mathrm{tr}\Big(\big(B_{t,\gamma_k}+V_{0,\gamma_k}+(\pi_{\gamma_k}^\circ-\pi_{0,\gamma_k|t}^\circ)\Lambda_{0,\gamma_k|t}^{-1}(\pi_{\gamma_k}^\circ-\pi_{0,\gamma_k|t}^\circ)'\big)\Sigma_{\alpha_k}^{-1}\Big)\bigg\},\nonumber
\end{eqnarray}
where the normalizing coefficient equals
\begin{equation}\label{08063}
c_4:=\frac{1}{(2\pi)^{ndr_{\gamma}/2}\prod_{k=1}^{r_{\gamma}}|\Lambda_{0,\gamma_k|t}|^{n/2}}\prod_{k=1}^{r_{\gamma}}\frac{\big|B_{t,\gamma_k}+V_{0,\gamma_k}\big|^{\nu_{0,\gamma_k|t}/2}}{\Gamma_{n}\big(\nu_{0,\gamma_k|t}/2\big)2^{n\nu_{0,\gamma_k|t}/2}}.
\end{equation}
If we integrate the above equation by $\Sigma_\gamma$, then one obtains that
\begin{eqnarray}\label{08064}
f(\pi_{\gamma}^\circ|\bar{s}_t,\mathcal{F}_t)&=&\prod_{k=1}^{r_{\gamma}}\frac{|\Lambda_{0,\gamma_k|t}|^{n/2}\big|B_{t,\gamma_k}+V_{0,\gamma_k}\big|^{-d/2}\Gamma_{n}\big((\nu_{0,\gamma_k|t}+d)/2\big)}{\pi^{nd/2}\Gamma_{n}(\nu_{0,\gamma_k|t}/2)}\\
&\times&\big|I_n+(B_{t,\gamma_k}+V_{0,\gamma_k})^{-1}(\pi_{\gamma_k}^\circ-\pi_{0,\gamma_k|t}^\circ)\Lambda_{0,\gamma_k|t}^{-1}(\pi_{\gamma_k}^\circ-\pi_{0,\gamma_k|t}^\circ)'\big|^{-(\nu_{0,\gamma_k|t}+d)/2}.\nonumber
\end{eqnarray}
Similarly, if $r_\delta>0$, it can be shown that
\begin{eqnarray}\label{08065}
f(\pi_{\delta}^\circ|\delta,\mathcal{F}_0)&=&\prod_{\ell=1}^{r_{\delta}}\frac{|\Lambda_{0,\delta_\ell}|^{-n/2}|V_{0,\delta_\ell}|^{-d/2}\Gamma_{n}\big((\nu_{0,\delta_\ell}+d)/2\big)}{\pi^{nd/2}\Gamma_{n}(\nu_{0,\delta_\ell}/2)}\nonumber\\
&\times&\big|I_n+V_{0,\delta_\ell}^{-1}(\pi_{\delta_\ell}^\circ-\pi_{0,\delta_\ell}^\circ)\Lambda_{0,\delta_\ell}^{-1}(\pi_{\delta_\ell}^\circ-\pi_{0,\delta_\ell}^\circ)'\big|^{-(\nu_{0,\delta_\ell}+d)/2}.
\end{eqnarray}
Therefore, equation \eqref{08047} holds. By the completing square method, the matrix $B_{t,\alpha_k}$ can be written by 
\begin{eqnarray}\label{08066}
B_{t,\alpha_k}&=&\big(y_{t,\alpha_k}^\circ-\pi_{0,\alpha_k}^\circ\Lambda_{0,\alpha_k}^{-1}\Lambda_{0,\alpha_k|t}\mathsf{Y}_{t,\alpha_k}^{\circ}\Phi_{t,\alpha_k}^{-1}\big)\Phi_{t,\alpha_k}^{-1}\big(y_{t,\alpha_k}^\circ-\pi_{0,\alpha_k}^\circ\Lambda_{0,\alpha_k}^{-1}\Lambda_{0,\alpha_k|t}\mathsf{Y}_{t,\alpha_k}^{\circ}\Phi_{t,\alpha_k}^{-1}\big)'\nonumber\\
&-&\pi_{0,\alpha_k}^\circ\Lambda_{0,\alpha_k}^{-1}\Lambda_{0,\alpha_k|t}\mathsf{Y}_{t,\alpha_k}^{\circ}\Phi_{t,\alpha_k}^{-1}(\mathsf{Y}_{t,\alpha_k}^{\circ})'\Lambda_{0,\alpha_k|t}\Lambda_{0,\alpha_k}^{-1}(\pi_{0,\alpha_k}^\circ)'\nonumber\\
&+&\pi_{0,\alpha_k}^\circ\Lambda_{0,\alpha_k}^{-1}(\pi_{0,\alpha_k}^\circ)'-\pi_{0,\alpha_k}^\circ\Lambda_{0,\alpha_k}^{-1}\Lambda_{0,\alpha_k|t}\Lambda_{0,\alpha_k}^{-1}(\pi_{0,\alpha_k}^\circ)',
\end{eqnarray}
where $\Phi_{t,\alpha_k}:=I_{q_{t,\alpha_k}}-(\mathsf{Y}_{t,\alpha_k}^\circ)'\Lambda_{0,\alpha_k|t}\mathsf{Y}_{t,\alpha_k}^\circ$ is a symmetric $(q_{t,\alpha_k}\times q_{t,\alpha_k})$ matrix. We consider the following product
\begin{equation}\label{08133}
I_{t,\alpha_k}:=\big(I_{q_{t,\alpha_k}}+(\mathsf{Y}_{t,\alpha_k}^\circ)'\Lambda_{0,\alpha_k}\mathsf{Y}_{t,\alpha_k}^\circ\big)\big(I_{q_{t,\alpha_k}}-(\mathsf{Y}_{t,\alpha_k}^\circ)'\Lambda_{0,\alpha_k|t}\mathsf{Y}_{t,\alpha_k}^\circ\big).
\end{equation}
It equals
\begin{eqnarray}\label{08134}
I_{t,\alpha_k}&=&I_{q_{t,\alpha_k}}+(\mathsf{Y}_{t,\alpha_k}^\circ)'\Lambda_{0,\alpha_k}\mathsf{Y}_{t,\alpha_k}^\circ-(\mathsf{Y}_{t,\alpha_k}^\circ)'\Lambda_{0,\alpha_k|t}\mathsf{Y}_{t,\alpha_k}^\circ\nonumber\\
&-&(\mathsf{Y}_{t,\alpha_k}^\circ)'\Lambda_{0,\alpha_k}\mathsf{Y}_{t,\alpha_k}^\circ(\mathsf{Y}_{t,\alpha_k}^\circ)'\Lambda_{0,\alpha_k|t}\mathsf{Y}_{t,\alpha_k}^\circ.
\end{eqnarray}
If we add and subtract the matrix $\Lambda_{0,\alpha_k}^{-1}$ into the term $\mathsf{Y}_{t,\alpha_k}^\circ(\mathsf{Y}_{t,\alpha_k}^\circ)'$ in the last line of the above equation, then the product matrix equals $I_{t,\alpha_k}=I_{q_{t,\alpha_k}}$. Consequently, the matrix $I_{q_{t,\alpha_k}}+(\mathsf{Y}_{t,\alpha_k}^\circ)'\Lambda_{0,\alpha_k}\mathsf{Y}_{t,\alpha_k}^\circ$ is an inverse matrix of the matrix $\Phi_{t,\alpha_k}$, that is,
\begin{equation}\label{08135}
\Phi_{t,\alpha_k}^{-1}=I_{q_{t,\alpha_k}}+(\mathsf{Y}_{t,\alpha_k}^\circ)'\Lambda_{0,\alpha_k}\mathsf{Y}_{t,\alpha_k}^\circ.
\end{equation}
Since it is a positive definite matrix, the matrix $B_{t,\alpha_k}$ is a positive semi--definite matrix. Now, we consider the term $\Lambda_{0,\alpha_k}^{-1}\Lambda_{0,\alpha_k|t}\mathsf{Y}_{t,\alpha_k}^{\circ}\Phi_{t,\alpha_k}^{-1}$ in the first line in equation \eqref{08066}. Similarly as before, by adding and subtracting $\Lambda_{0,\alpha_k}^{-1}$ into the term $\mathsf{Y}_{t,\alpha_k}^\circ(\mathsf{Y}_{t,\alpha_k}^\circ)'$, one obtains that
\begin{equation}\label{08139}
\Lambda_{0,\alpha_k}^{-1}\Lambda_{0,\alpha_k|t}\mathsf{Y}_{t,\alpha_k}^{\circ}\Phi_{t,\alpha_k}^{-1}=\mathsf{Y}_{t,\alpha_k}^{\circ}.
\end{equation}
Consequently, the sum of the second and third lines of equation \eqref{08066} equals zero. Let $\Lambda_{0,\alpha_k}^{1/2}$ be the Cholesky factor of the matrix $\Lambda_{0,\alpha_k}$, i.e., $\Lambda_{0,\alpha_k}=\big(\Lambda_{0,\alpha_k}^{1/2}\big)'\Lambda_{0,\alpha_k}^{1/2}$. Then, according to the Sylvester's determinant theorem, see \citeA{Lutkepohl05}, a determinant of the matrix $\Phi_{t,\alpha_k}^{-1}$ is
\begin{equation}\label{08140}
|\Phi_{t,\alpha_k}^{-1}|=\big|I_{q_{t,\alpha_k}}+\big(\Lambda_{0,s_t}^{1/2}\mathsf{Y}_{t,\alpha_k}^\circ\big)'\Lambda_{0,s_t}^{1/2}\mathsf{Y}_{t,\alpha_k}^\circ\big|=\big|I_d+\Lambda_{0,s_t}^{1/2}\mathsf{Y}_{t,\alpha_k}^\circ\big(\mathsf{Y}_{t,\alpha_k}^\circ\big)'\big(\Lambda_{0,s_t}^{1/2}\big)'\big|.
\end{equation}
Finally, let us assume $q_{t,\alpha_k}^*>0$. Since $\mathsf{Y}_{T,\alpha_k}^\circ=[\mathsf{Y}_{t,\alpha_k}^\circ:\mathsf{Y}_{t,\alpha_k}^*]$, we have that
\begin{equation}\label{•}
I_{q_{T,\alpha_k}}+(\mathsf{Y}_{T,\alpha_k}^\circ)'\Lambda_{0,\alpha_k}\mathsf{Y}_{T,\alpha_k}^\circ=\begin{bmatrix}
C_{11} & C_{12}\\
C_{21} & C_{22}
\end{bmatrix},
\end{equation}
where $C_{11}:=I_{q_{t,\alpha_k}}+(\mathsf{Y}_{t,\alpha_k}^\circ)'\Lambda_{0,\alpha_k}\mathsf{Y}_{t,\alpha_k}^\circ$, $C_{12}:=(\mathsf{Y}_{t,\alpha_k}^\circ)'\Lambda_{0,\alpha_k}\mathsf{Y}_{t,\alpha_k}^*$, $C_{21}:=(\mathsf{Y}_{t,\alpha_k}^*)'\Lambda_{0,\alpha_k}\mathsf{Y}_{t,\alpha_k}^\circ$, and $C_{22}:=I_{q_{t,\alpha_k}^*}+(\mathsf{Y}_{t,\alpha_k}^*)'\Lambda_{0,\alpha_k}\mathsf{Y}_{t,\alpha_k}^*$.
Due to the well--known formula of inverse of partitioned matrix, inverse of the above matrix is 
\begin{equation}\label{•}
\big(I_{q_{T,\alpha_k}}+(\mathsf{Y}_{T,\alpha_k}^\circ)'\Lambda_{0,\alpha_k}\mathsf{Y}_{T,\alpha_k}^\circ\big)^{-1}=\begin{bmatrix}
C_{11}^{-1}+C_{11}^{-1}C_{12}C_{22}^{-1}C_{21}C_{11}^{-1} & -C_{11}^{-1}C_{12}D_{22}\\
-D_{22}C_{21}C_{11}^{-1} & D_{22}
\end{bmatrix}.
\end{equation}
where $D_{22}:=(C_{22}-C_{21}C_{11}^{-1}C_{12})^{-1}$. To simplify notations, we define $\Delta_1:=y_{t,\alpha_k}^\circ-\pi_{0,\alpha_k}^\circ \mathsf{Y}_{t,\alpha_k}^\circ$ and $\Delta_2:=y_{t,\alpha_k}^*-\pi_{0,\alpha_k}^\circ \mathsf{Y}_{t,\alpha_k}^*$. Consequently, the matrix $B_{T,\alpha_k}$ is represented by
\begin{eqnarray}\label{•}
B_{T,\alpha_k}&=&\begin{bmatrix}
\Delta_1 & \Delta_2
\end{bmatrix}
\begin{bmatrix}
C_{11}^{-1}+C_{11}^{-1}C_{12}C_{22}^{-1}C_{21}C_{11}^{-1} & -C_{11}^{-1}C_{12}D_{22}\\
-D_{22}C_{21}C_{11}^{-1} & D_{22}
\end{bmatrix}
\begin{bmatrix}
\Delta_1' \\ \Delta_2'
\end{bmatrix}\\
&=&B_{t,\alpha_k}+(\Delta_2-\Delta_1C_{11}^{-1}C_{12})D_{22}(\Delta_2-\Delta_1C_{11}^{-1}C_{12})'.
\end{eqnarray}
Also, the matrix $D_{22}$ equals
\begin{equation}\label{08160}
D_{22}=\big(I_{q_{t,\alpha_k}^*}+(\mathsf{Y}_{t,\alpha_k}^*)'\Theta_{t,\alpha_k}\mathsf{Y}_{t,\alpha_k}^*\big)^{-1},
\end{equation}
where $\Theta_{t,\alpha_k}:=\Lambda_{0,\alpha_k}-\Lambda_{0,\alpha_k}\mathsf{Y}_{t,\alpha_k}^\circ\Phi_{t,\alpha_k}(\mathsf{Y}_{t,\alpha_k}^\circ)'\Lambda_{0,\alpha_k}.$
Since $\Phi_{t,\alpha_k}=I_{q_{t,\alpha_k}}-(\mathsf{Y}_{t,\alpha_k}^\circ)'\Lambda_{0,\alpha_k|t}\mathsf{Y}_{t,\alpha_k}^\circ$, we have that
\begin{equation}\label{08161}
\Theta_{t,\alpha_k}=\Lambda_{0,\alpha_k}-\Lambda_{0,\alpha_k}\mathsf{Y}_{t,\alpha_k}^\circ(\mathsf{Y}_{t,\alpha_k}^\circ)'\Lambda_{0,\alpha_k}+\Lambda_{0,\alpha_k}\mathsf{Y}_{t,\alpha_k}^\circ(\mathsf{Y}_{t,\alpha_k}^\circ)'\Lambda_{0,\alpha_k|t}\mathsf{Y}_{t,\alpha_k}^\circ(\mathsf{Y}_{t,\alpha_k}^\circ)'\Lambda_{0,\alpha_k}.
\end{equation}
If we substitute the matrix $\mathsf{Y}_{t,\alpha_k}^\circ(\mathsf{Y}_{t,\alpha_k}^\circ)'=\Lambda_{0,\alpha_k|t}^{-1}-\Lambda_{0,\alpha_k}^{-1}$ into the above equation, one obtains $\Theta_{t,\alpha_k}=\Lambda_{0,\alpha_k|t}$. As a result, equation \eqref{08162} holds. That completes the proof of the Proposition.
\end{proof}

\begin{proof}[\textbf{Proof of Proposition \ref{prop03}}]
Since for $t=2,\dots,T$, the random variable $n_{ij}(\bar{s}_t)$ represents a number of consequential elements, which equals $(i,j)$ of the regime vector $\bar{s}_t$, we have that 
\begin{equation}\label{08079}
f(\bar{s}_t|\mathsf{P},\mathcal{F}_0)=\prod_{m=1}^tp_{s_{m-1}s_m}=\prod_{i=0}^N\prod_{j=1}^Np_{ij}^{n_{ij}(\bar{s}_t)}.
\end{equation}
Consequently, it follows from the joint density function of the random transition probability matrix $\mathsf{P}$, which is given by equation \eqref{08069} that
\begin{equation}\label{08080}
f(\bar{s}_t|\mathcal{F}_0)=\int_{\mathsf{P}}f(\bar{s}_t|\mathsf{P},\mathcal{F}_0)f(\mathsf{P}|\mathcal{F}_0)d\mathsf{P}=\prod_{i=0}^N\frac{\Gamma\big(\sum_{j=1}^N\alpha_{ij}\big)}{\prod_{j=1}^N\Gamma(\alpha_{ij})}\frac{\prod_{j=1}^N\Gamma(\alpha_{ij}+n_{ij}(\bar{s}_t))}{\Gamma\big(\sum_{j=1}^N(\alpha_{ij}+n_{ij}(\bar{s}_t))\big)}.
\end{equation}
On the other hand, as $f(s_t|\bar{s}_{t},\mathcal{F}_0)=f(\bar{s}_t|\mathcal{F}_0)/f(\bar{s}_{t-1}|\mathcal{F}_0)$, one obtains 
\begin{equation}\label{08081}
f(s_t|\bar{s}_{t-1},\mathcal{F}_0)=\prod_{i=0}^N\frac{\prod_{j=1}^N\Gamma(\alpha_{ij}+n_{ij}(\bar{s}_t))}{\Gamma\big(\sum_{j=1}^N(\alpha_{ij}+n_{ij}(\bar{s}_t))\big)}\frac{\Gamma\big(\sum_{j=1}^N(\alpha_{ij}+n_{ij}(\bar{s}_{t-1}))\big)}{\prod_{j=1}^N\Gamma(\alpha_{ij}+n_{ij}(\bar{s}_{t-1}))}.
\end{equation}
Consequently, since $n_{ij}(\bar{s}_t)=n_{ij}(\bar{s}_{t-1})+\delta_{ij}(s_t)$ for $t=2,\dots,T$ and $\Gamma(x+1)=x\Gamma(x)$ for $x>0$, we find that
\begin{equation}\label{08082}
f(s_t|\bar{s}_{t-1},\mathcal{F}_0)=\frac{\alpha_{s_{t-1}s_t}+n_{s_{t-1}s_t}(\bar{s}_{t-1})}{\sum_{s_t=1}^N\big(\alpha_{s_{t-1}s_t}+n_{s_{t-1}s_t}(\bar{s}_{t-1})\big)},
\end{equation}
where the random variable $\delta_{ij}(s_t)$ equals
\begin{equation}\label{08083}
\delta_{ij}(s_t)=\begin{cases}
1 & \mathrm{if}~~~s_{t-1}=i,~s_t=j,\\
0 & \mathrm{if}~~~\mathrm{otherwise}.
\end{cases}
\end{equation}
That completes the proof. 
\end{proof}

\begin{proof}[\textbf{Proof of Proposition \ref{prop04}}]
The proof omits the iteration number $\ell$. According to the conditional probability formula, we have that
\begin{equation}\label{08171}
f(\bar{s}_t|\lambda_t,\mathcal{F}_t)=\frac{f(y_t|\bar{s}_t,\lambda_t,\mathcal{F}_{t-1})f(s_t|\bar{s}_{t-1},\lambda_t,\mathcal{F}_{t-1})f(\bar{y}_{t-1},\bar{s}_{t-1}|\lambda_t,\mathcal{F}_0)}{f(\bar{y}_t|\lambda_t,\mathcal{F}_0)}.
\end{equation}
The first density function in numerator equals $\eta_{t,s_t}$. For the second density function, we have that
\begin{equation}\label{08172}
f(s_t|\bar{s}_{t-1},\lambda_t,\mathcal{F}_{t-1})=\frac{f(\bar{s}_t,\lambda_t|\mathcal{F}_{t-1})}{\sum_{s_t}f(\bar{s}_t,\lambda_t|\mathcal{F}_{t-1})}=\frac{f_*(\pi_{s_t},\Sigma_{s_t}|\bar{s}_t,\lambda_{t-1},\mathcal{F}_{t-1})p_{s_{t-1}s_t}}{\sum_{s_t}f_*(\pi_{s_t},\Sigma_{s_t}|\bar{s}_t,\lambda_{t-1},\mathcal{F}_{t-1})p_{s_{t-1}s_t}},
\end{equation}
where the parameter $\lambda_{t-1}$ corresponds to the regime vector $\bar{s}_{t-1}$ and we use a fact that due to Markov property \eqref{08156}, $f(\bar{s}_t,\lambda_t|\mathcal{F}_{t-1})=f_*(\pi_{s_t},\Sigma_{s_t}|\bar{s}_t,\lambda_{t-1},\mathcal{F}_{t-1})p_{s_{t-1}s_t}f(\bar{s}_{t-1}|\lambda_{t-1},\mathcal{F}_{t-1})$. As a result, according to the assumption, the second density function is approximated by $p_{s_{t-1}s_t}$. By the conditional probability formula, the denominator of equation \eqref{08171} equals $f(y_t|\lambda_t,\mathcal{F}_{t-1})f(\bar{y}_{t-1}|\lambda_t,\mathcal{F}_0)$. Consequently, since $f(y_t|\lambda_t,\mathcal{F}_{t-1})=i_N'(z_{t|t-1}\odot\eta_t)$ (see \citeA{Hamilton94}), one gets that
\begin{equation}\label{08173}
f(\bar{s}_t|\lambda_t,\mathcal{F}_t)=\frac{\eta_{t,s_t}p_{s_{t-1}s_t}}{i_N'(z_{t|t-1}\odot\eta_t)}\times f(\bar{s}_{t-1}|\lambda_t,\mathcal{F}_{t-1}).
\end{equation}
If we repeat the above equation, then for $u=1,\dots,t-1$, one obtains that
\begin{equation}\label{08174}
f(\bar{s}_t|\lambda_t,\mathcal{F}_t)=\frac{\eta_{u+1,s_{u+1}}p_{s_us_{u+1}}\dots \eta_{t,s_t}p_{s_{t-1}s_t}}{i_N'(z_{u+1|u}\odot\eta_{u+1})\dots i_N'(z_{t|t-1}\odot\eta_t)}\times f(\bar{s}_u|\lambda_t,\mathcal{F}_u).
\end{equation}
If we integrate the above equation by $\bar{s}_{u:t}^c:=(s_{u+1},\dots,s_t)'$ and $\bar{s}_{u-1}$, then as the denominator does not depend on the regime vector $s$ we get that
\begin{equation}\label{08175}
f(s_u|\lambda_t,\mathcal{F}_t)=\frac{\sum_{\bar{s}_{u:t}^c}\eta_{u+1,s_{u+1}}p_{s_us_{u+1}}\dots \eta_{t,s_t}p_{s_{t-1}s_t}}{i_N'(z_{u+1|u}\odot\eta_{u+1})\dots i_N'(z_{t|t-1}\odot\eta_t)}\times f(s_u|\lambda_t,\mathcal{F}_u).
\end{equation}
Similarly, one obtains that
\begin{equation}\label{08176}
f(s_{u-1},s_u|\lambda_t,\mathcal{F}_t)=\frac{\sum_{\bar{s}_{u:t}^c}\eta_{u+1,s_{u+1}}p_{s_us_{u+1}}\dots \eta_{t,s_t}p_{s_{t-1}s_t}}{i_N'(z_{u+1|u}\odot\eta_{u+1})\dots i_N'(z_{t|t-1}\odot\eta_t)}\times f(s_{u-1},s_u|\lambda_t,\mathcal{F}_u),
\end{equation}
where by equation \eqref{08173}, the density function $f(s_{u-1},s_u|\lambda_t,\mathcal{F}_u)$ equals
\begin{equation}\label{08177}
f(s_{u-1},s_u|\lambda_t,\mathcal{F}_u)=\frac{\eta_{u,s_u}p_{s_{u-1}s_u}}{i_N'(z_{u|u-1}\odot\eta_u)}\times f(s_{u-1}|\lambda_t,\mathcal{F}_{u-1}).
\end{equation}
Consequently, we have that
\begin{equation}\label{08178}
z_{u|t}=\frac{\big(\hat{\mathsf{P}}\mathsf{H}_{u+1}\dots \hat{\mathsf{P}}\mathsf{H}_ti_N\big)\odot z_{u|u}}{i_N'(z_{u+1|u}\odot\eta_{u+1})\dots i_N'(z_{t|t-1}\odot\eta_t)}, ~~~u=t-1,\dots,1
\end{equation}
and
\begin{equation}\label{08179}
f(s_{u-1},s_u|\lambda_t,\mathcal{F}_t)=\frac{\big(z_{u|t}\oslash z_{u|u}\big)_{s_u}\eta_{u,s_u}p_{s_{u-1}s_u}\big(z_{u-1|u-1}\big)_{s_{u-1}}}{i_N'(z_{u|u-1}\odot\eta_u)},~~~u=2,\dots,t,
\end{equation}
for the products of the matrices in $z_{u|t}$, see \citeA{Zucchini16}. As a result, one obtains equations \eqref{08110}, \eqref{08180}, and \eqref{08165}. That completes the proof.
\end{proof}

\bibliographystyle{apacite}
\bibliography{References}

\begin{thebibliography}{}

\bibitem [\protect \citeauthoryear {%
Albert%
\ \BBA {} Chib%
}{%
Albert%
\ \BBA {} Chib%
}{%
{\protect \APACyear {1993}}%
}]{%
Albert93}
\APACinsertmetastar {%
Albert93}%
\begin{APACrefauthors}%
Albert, J\BPBI H.%
\BCBT {}\ \BBA {} Chib, S.%
\end{APACrefauthors}%
\unskip\
\newblock
\APACrefYearMonthDay{1993}{}{}.
\newblock
{\BBOQ}\APACrefatitle {Bayes Inference via Gibbs Sampling of Autoregressive
  Time Series Subject to Markov Mean and Variance Shifts} {Bayes inference via
  gibbs sampling of autoregressive time series subject to markov mean and
  variance shifts}.{\BBCQ}
\newblock
\APACjournalVolNumPages{Journal of Business \& Economic
  Statistics}{11}{1}{1--15}.
\newblock
\begin{APACrefDOI} \doi{10.2307/1391303} \end{APACrefDOI}
\PrintBackRefs{\CurrentBib}

\bibitem [\protect \citeauthoryear {%
Ba{\'n}bura%
, Giannone%
\BCBL {}\ \BBA {} Reichlin%
}{%
Ba{\'n}bura%
\ \protect \BOthers {.}}{%
{\protect \APACyear {2010}}%
}]{%
Banbura10}
\APACinsertmetastar {%
Banbura10}%
\begin{APACrefauthors}%
Ba{\'n}bura, M.%
, Giannone, D.%
\BCBL {}\ \BBA {} Reichlin, L.%
\end{APACrefauthors}%
\unskip\
\newblock
\APACrefYearMonthDay{2010}{}{}.
\newblock
{\BBOQ}\APACrefatitle {{Large Bayesian Vector Autoregressions}} {{Large
  Bayesian Vector Autoregressions}}.{\BBCQ}
\newblock
\APACjournalVolNumPages{Journal of Applied Econometrics}{25}{1}{71--92}.
\PrintBackRefs{\CurrentBib}

\bibitem [\protect \citeauthoryear {%
Battulga%
}{%
Battulga%
}{%
{\protect \APACyear {2023}}%
{\protect \APACexlab {{\protect \BCnt {1}}}}}]{%
Battulga23b}
\APACinsertmetastar {%
Battulga23b}%
\begin{APACrefauthors}%
Battulga, G.%
\end{APACrefauthors}%
\unskip\
\newblock
\APACrefYearMonthDay{2023{\protect \BCnt {1}}}{}{}.
\newblock
{\BBOQ}\APACrefatitle {{Parameter Estimation Methods of Required Rate of Return
  on Stock}} {{Parameter Estimation Methods of Required Rate of Return on
  Stock}}.{\BBCQ}
\newblock
\APACjournalVolNumPages{International Journal of Theoretical and Applied
  Finance}{26}{8}{2450005}.
\PrintBackRefs{\CurrentBib}

\bibitem [\protect \citeauthoryear {%
Battulga%
}{%
Battulga%
}{%
{\protect \APACyear {2023}}%
{\protect \APACexlab {{\protect \BCnt {2}}}}}]{%
Battulga23a}
\APACinsertmetastar {%
Battulga23a}%
\begin{APACrefauthors}%
Battulga, G.%
\end{APACrefauthors}%
\unskip\
\newblock
\APACrefYearMonthDay{2023{\protect \BCnt {2}}}{}{}.
\newblock
{\BBOQ}\APACrefatitle {{Rainbow Options with Bayesian MS-VAR Process}}
  {{Rainbow Options with Bayesian MS-VAR Process}}.{\BBCQ}
\newblock
\APACjournalVolNumPages{Mongolian Mathematical Journal}{26}{24}{1-16}.
\PrintBackRefs{\CurrentBib}

\bibitem [\protect \citeauthoryear {%
Battulga%
}{%
Battulga%
}{%
{\protect \APACyear {2024}}%
{\protect \APACexlab {{\protect \BCnt {1}}}}}]{%
Battulga24f}
\APACinsertmetastar {%
Battulga24f}%
\begin{APACrefauthors}%
Battulga, G.%
\end{APACrefauthors}%
\unskip\
\newblock
\APACrefYearMonthDay{2024{\protect \BCnt {1}}}{}{}.
\newblock
{\BBOQ}\APACrefatitle {{Equity--Linked Life Insurances on Maximum of Several
  Assets}} {{Equity--Linked Life Insurances on Maximum of Several
  Assets}}.{\BBCQ}
\newblock
\APACjournalVolNumPages{to appear in Numerical Algebra, Control \&
  Optimization}{}{}{}.
\newblock
\APAChowpublished {Available at: \url{https://arxiv.org/abs/2112.10447}}.
\PrintBackRefs{\CurrentBib}

\bibitem [\protect \citeauthoryear {%
Battulga%
}{%
Battulga%
}{%
{\protect \APACyear {2024}}%
{\protect \APACexlab {{\protect \BCnt {2}}}}}]{%
Battulga24a}
\APACinsertmetastar {%
Battulga24a}%
\begin{APACrefauthors}%
Battulga, G.%
\end{APACrefauthors}%
\unskip\
\newblock
\APACrefYearMonthDay{2024{\protect \BCnt {2}}}{}{}.
\newblock
{\BBOQ}\APACrefatitle {{Options Pricing under Bayesian MS--VAR Process}}
  {{Options Pricing under Bayesian MS--VAR Process}}.{\BBCQ}
\newblock
\APACjournalVolNumPages{to appear in Numerical Algebra, Control \&
  Optimization}{}{}{}.
\newblock
\APAChowpublished {Available at: \url{https://arxiv.org/abs/2109.05998}}.
\PrintBackRefs{\CurrentBib}

\bibitem [\protect \citeauthoryear {%
Battulga%
}{%
Battulga%
}{%
{\protect \APACyear {2024}}%
{\protect \APACexlab {{\protect \BCnt {3}}}}}]{%
Battulga22b}
\APACinsertmetastar {%
Battulga22b}%
\begin{APACrefauthors}%
Battulga, G.%
\end{APACrefauthors}%
\unskip\
\newblock
\APACrefYearMonthDay{2024{\protect \BCnt {3}}}{}{}.
\newblock
{\BBOQ}\APACrefatitle {Stochastic DDM with regime--switching process}
  {Stochastic ddm with regime--switching process}.{\BBCQ}
\newblock
\APACjournalVolNumPages{Numerical Algebra, Control and
  Optimization}{14}{2}{339--365}.
\PrintBackRefs{\CurrentBib}

\bibitem [\protect \citeauthoryear {%
Butler%
}{%
Butler%
}{%
{\protect \APACyear {1998}}%
}]{%
Butler98}
\APACinsertmetastar {%
Butler98}%
\begin{APACrefauthors}%
Butler, R\BPBI W.%
\end{APACrefauthors}%
\unskip\
\newblock
\APACrefYearMonthDay{1998}{}{}.
\newblock
{\BBOQ}\APACrefatitle {Generalized inverse Gaussian distributions and their
  Wishart connections} {Generalized inverse gaussian distributions and their
  wishart connections}.{\BBCQ}
\newblock
\APACjournalVolNumPages{Scandinavian journal of statistics}{25}{1}{69--75}.
\PrintBackRefs{\CurrentBib}

\bibitem [\protect \citeauthoryear {%
Glasserman%
, Heidelberger%
\BCBL {}\ \BBA {} Shahabuddin%
}{%
Glasserman%
\ \protect \BOthers {.}}{%
{\protect \APACyear {2000}}%
}]{%
Glasserman00}
\APACinsertmetastar {%
Glasserman00}%
\begin{APACrefauthors}%
Glasserman, P.%
, Heidelberger, P.%
\BCBL {}\ \BBA {} Shahabuddin, P.%
\end{APACrefauthors}%
\unskip\
\newblock
\APACrefYearMonthDay{2000}{}{}.
\newblock
{\BBOQ}\APACrefatitle {Variance reduction techniques for estimating
  value-at-risk} {Variance reduction techniques for estimating
  value-at-risk}.{\BBCQ}
\newblock
\APACjournalVolNumPages{Management Science}{46}{10}{1349--1364}.
\PrintBackRefs{\CurrentBib}

\bibitem [\protect \citeauthoryear {%
Glasserman%
\ \BBA {} Li%
}{%
Glasserman%
\ \BBA {} Li%
}{%
{\protect \APACyear {2005}}%
}]{%
Glasserman05b}
\APACinsertmetastar {%
Glasserman05b}%
\begin{APACrefauthors}%
Glasserman, P.%
\BCBT {}\ \BBA {} Li, J.%
\end{APACrefauthors}%
\unskip\
\newblock
\APACrefYearMonthDay{2005}{}{}.
\newblock
{\BBOQ}\APACrefatitle {Importance sampling for portfolio credit risk}
  {Importance sampling for portfolio credit risk}.{\BBCQ}
\newblock
\APACjournalVolNumPages{Management science}{51}{11}{1643--1656}.
\PrintBackRefs{\CurrentBib}

\bibitem [\protect \citeauthoryear {%
Goldfeld%
\ \BBA {} Quandt%
}{%
Goldfeld%
\ \BBA {} Quandt%
}{%
{\protect \APACyear {1973}}%
}]{%
Goldfeld73}
\APACinsertmetastar {%
Goldfeld73}%
\begin{APACrefauthors}%
Goldfeld, S\BPBI M.%
\BCBT {}\ \BBA {} Quandt, R\BPBI E.%
\end{APACrefauthors}%
\unskip\
\newblock
\APACrefYearMonthDay{1973}{}{}.
\newblock
{\BBOQ}\APACrefatitle {A Markov model for switching regressions} {A markov
  model for switching regressions}.{\BBCQ}
\newblock
\APACjournalVolNumPages{Journal of Econometrics}{1}{1}{3--15}.
\PrintBackRefs{\CurrentBib}

\bibitem [\protect \citeauthoryear {%
Hamilton%
}{%
Hamilton%
}{%
{\protect \APACyear {1989}}%
}]{%
Hamilton89}
\APACinsertmetastar {%
Hamilton89}%
\begin{APACrefauthors}%
Hamilton, J\BPBI D.%
\end{APACrefauthors}%
\unskip\
\newblock
\APACrefYearMonthDay{1989}{}{}.
\newblock
{\BBOQ}\APACrefatitle {{A New Approach to the Economic Analysis of
  Nonstationary Time Series and the Business Cycle}} {{A New Approach to the
  Economic Analysis of Nonstationary Time Series and the Business
  Cycle}}.{\BBCQ}
\newblock
\APACjournalVolNumPages{Econometrica: Journal of the Econometric
  Society}{}{}{357--384}.
\PrintBackRefs{\CurrentBib}

\bibitem [\protect \citeauthoryear {%
Hamilton%
}{%
Hamilton%
}{%
{\protect \APACyear {1990}}%
}]{%
Hamilton90}
\APACinsertmetastar {%
Hamilton90}%
\begin{APACrefauthors}%
Hamilton, J\BPBI D.%
\end{APACrefauthors}%
\unskip\
\newblock
\APACrefYearMonthDay{1990}{}{}.
\newblock
{\BBOQ}\APACrefatitle {{Analysis of Time Series Subject to Changes in Regime}}
  {{Analysis of Time Series Subject to Changes in Regime}}.{\BBCQ}
\newblock
\APACjournalVolNumPages{Journal of Econometrics}{45}{1-2}{39--70}.
\PrintBackRefs{\CurrentBib}

\bibitem [\protect \citeauthoryear {%
Hamilton%
}{%
Hamilton%
}{%
{\protect \APACyear {1994}}%
}]{%
Hamilton94}
\APACinsertmetastar {%
Hamilton94}%
\begin{APACrefauthors}%
Hamilton, J\BPBI D.%
\end{APACrefauthors}%
\unskip\
\newblock
\APACrefYear{1994}.
\newblock
\APACrefbtitle {{Time Series Econometrics}} {{Time Series Econometrics}}.
\newblock
\APACaddressPublisher{}{Princeton University Press, Princeton}.
\PrintBackRefs{\CurrentBib}

\bibitem [\protect \citeauthoryear {%
Karlsson%
}{%
Karlsson%
}{%
{\protect \APACyear {2013}}%
}]{%
Karlsson13}
\APACinsertmetastar {%
Karlsson13}%
\begin{APACrefauthors}%
Karlsson, S.%
\end{APACrefauthors}%
\unskip\
\newblock
\APACrefYearMonthDay{2013}{}{}.
\newblock
{\BBOQ}\APACrefatitle {Forecasting with Bayesian vector autoregression}
  {Forecasting with bayesian vector autoregression}.{\BBCQ}
\newblock
\APACjournalVolNumPages{Handbook of economic forecasting}{2}{}{791--897}.
\PrintBackRefs{\CurrentBib}

\bibitem [\protect \citeauthoryear {%
Kim%
}{%
Kim%
}{%
{\protect \APACyear {1994}}%
}]{%
Kim94}
\APACinsertmetastar {%
Kim94}%
\begin{APACrefauthors}%
Kim, C\BHBI J.%
\end{APACrefauthors}%
\unskip\
\newblock
\APACrefYearMonthDay{1994}{}{}.
\newblock
{\BBOQ}\APACrefatitle {{Dynamic Linear Models with Markov--Switching}}
  {{Dynamic Linear Models with Markov--Switching}}.{\BBCQ}
\newblock
\APACjournalVolNumPages{Journal of Econometrics}{60}{1--2}{1--22}.
\PrintBackRefs{\CurrentBib}

\bibitem [\protect \citeauthoryear {%
Krolzig%
}{%
Krolzig%
}{%
{\protect \APACyear {1997}}%
}]{%
Krolzig97}
\APACinsertmetastar {%
Krolzig97}%
\begin{APACrefauthors}%
Krolzig, H\BHBI M.%
\end{APACrefauthors}%
\unskip\
\newblock
\APACrefYear{1997}.
\newblock
\APACrefbtitle {Markov-switching vector autoregressions: Modelling, statistical
  inference, and application to business cycle analysis} {Markov-switching
  vector autoregressions: Modelling, statistical inference, and application to
  business cycle analysis}\ (\BVOL~454).
\newblock
\APACaddressPublisher{}{Springer Science \& Business Media}.
\PrintBackRefs{\CurrentBib}

\bibitem [\protect \citeauthoryear {%
Litterman%
}{%
Litterman%
}{%
{\protect \APACyear {1979}}%
}]{%
Litterman79}
\APACinsertmetastar {%
Litterman79}%
\begin{APACrefauthors}%
Litterman, R.%
\end{APACrefauthors}%
\unskip\
\newblock
\APACrefYearMonthDay{1979}{}{}.
\newblock
{\BBOQ}\APACrefatitle {Techniques of forecasting using vector autoregressions}
  {Techniques of forecasting using vector autoregressions}.{\BBCQ}
\newblock
\APACjournalVolNumPages{Federal Reserve of Minneapolis Working Paper
  115}{}{}{}.
\PrintBackRefs{\CurrentBib}

\bibitem [\protect \citeauthoryear {%
L{\"u}tkepohl%
}{%
L{\"u}tkepohl%
}{%
{\protect \APACyear {2005}}%
}]{%
Lutkepohl05}
\APACinsertmetastar {%
Lutkepohl05}%
\begin{APACrefauthors}%
L{\"u}tkepohl, H.%
\end{APACrefauthors}%
\unskip\
\newblock
\APACrefYear{2005}.
\newblock
\APACrefbtitle {{New Introduction to Multiple Time Series Analysis}} {{New
  Introduction to Multiple Time Series Analysis}}\ (\PrintOrdinal{2}\ \BEd).
\newblock
\APACaddressPublisher{}{Springer Berlin Heidelberg}.
\PrintBackRefs{\CurrentBib}

\bibitem [\protect \citeauthoryear {%
McNeil%
, Frey%
\BCBL {}\ \BBA {} Embrechts%
}{%
McNeil%
\ \protect \BOthers {.}}{%
{\protect \APACyear {2005}}%
}]{%
McNeil05}
\APACinsertmetastar {%
McNeil05}%
\begin{APACrefauthors}%
McNeil, A\BPBI J.%
, Frey, R.%
\BCBL {}\ \BBA {} Embrechts, P.%
\end{APACrefauthors}%
\unskip\
\newblock
\APACrefYear{2005}.
\newblock
\APACrefbtitle {{Quantitative Risk Management: Concepts, Techniques and Tools}}
  {{Quantitative Risk Management: Concepts, Techniques and Tools}}.
\newblock
\APACaddressPublisher{}{Princeton University Press}.
\PrintBackRefs{\CurrentBib}

\bibitem [\protect \citeauthoryear {%
Miranda-Agrippino%
\ \BBA {} Ricco%
}{%
Miranda-Agrippino%
\ \BBA {} Ricco%
}{%
{\protect \APACyear {2018}}%
}]{%
Miranda18}
\APACinsertmetastar {%
Miranda18}%
\begin{APACrefauthors}%
Miranda-Agrippino, S.%
\BCBT {}\ \BBA {} Ricco, G.%
\end{APACrefauthors}%
\unskip\
\newblock
\APACrefYearMonthDay{2018}{}{}.
\newblock
{\BBOQ}\APACrefatitle {Bayesian vector autoregressions} {Bayesian vector
  autoregressions}.{\BBCQ}
\newblock

\PrintBackRefs{\CurrentBib}

\bibitem [\protect \citeauthoryear {%
Quandt%
}{%
Quandt%
}{%
{\protect \APACyear {1958}}%
}]{%
Quandt58}
\APACinsertmetastar {%
Quandt58}%
\begin{APACrefauthors}%
Quandt, R\BPBI E.%
\end{APACrefauthors}%
\unskip\
\newblock
\APACrefYearMonthDay{1958}{}{}.
\newblock
{\BBOQ}\APACrefatitle {The estimation of the parameters of a linear regression
  system obeying two separate regimes} {The estimation of the parameters of a
  linear regression system obeying two separate regimes}.{\BBCQ}
\newblock
\APACjournalVolNumPages{Journal of the american statistical
  association}{53}{284}{873--880}.
\PrintBackRefs{\CurrentBib}

\bibitem [\protect \citeauthoryear {%
Sims%
}{%
Sims%
}{%
{\protect \APACyear {1980}}%
}]{%
Sims80}
\APACinsertmetastar {%
Sims80}%
\begin{APACrefauthors}%
Sims, C\BPBI A.%
\end{APACrefauthors}%
\unskip\
\newblock
\APACrefYearMonthDay{1980}{}{}.
\newblock
{\BBOQ}\APACrefatitle {Macroeconomics and Reality} {Macroeconomics and
  reality}.{\BBCQ}
\newblock
\APACjournalVolNumPages{Econometrica}{48}{1}{1--48}.
\PrintBackRefs{\CurrentBib}

\bibitem [\protect \citeauthoryear {%
Tiao%
\ \BBA {} Box%
}{%
Tiao%
\ \BBA {} Box%
}{%
{\protect \APACyear {1981}}%
}]{%
Tiao81}
\APACinsertmetastar {%
Tiao81}%
\begin{APACrefauthors}%
Tiao, G\BPBI C.%
\BCBT {}\ \BBA {} Box, G\BPBI E.%
\end{APACrefauthors}%
\unskip\
\newblock
\APACrefYearMonthDay{1981}{}{}.
\newblock
{\BBOQ}\APACrefatitle {Modeling multiple time series with applications}
  {Modeling multiple time series with applications}.{\BBCQ}
\newblock
\APACjournalVolNumPages{Journal of the American Statistical
  Association}{76}{376}{802--816}.
\PrintBackRefs{\CurrentBib}

\bibitem [\protect \citeauthoryear {%
Tong%
}{%
Tong%
}{%
{\protect \APACyear {1983}}%
}]{%
Tong83}
\APACinsertmetastar {%
Tong83}%
\begin{APACrefauthors}%
Tong, H.%
\end{APACrefauthors}%
\unskip\
\newblock
\APACrefYear{1983}.
\newblock
\APACrefbtitle {Threshold models in non-linear time series analysis} {Threshold
  models in non-linear time series analysis}\ (\BVOL~21).
\newblock
\APACaddressPublisher{}{Springer Science \& Business Media}.
\PrintBackRefs{\CurrentBib}

\bibitem [\protect \citeauthoryear {%
Zucchini%
, MacDonald%
\BCBL {}\ \BBA {} Langrock%
}{%
Zucchini%
\ \protect \BOthers {.}}{%
{\protect \APACyear {2016}}%
}]{%
Zucchini16}
\APACinsertmetastar {%
Zucchini16}%
\begin{APACrefauthors}%
Zucchini, W.%
, MacDonald, I\BPBI L.%
\BCBL {}\ \BBA {} Langrock, R.%
\end{APACrefauthors}%
\unskip\
\newblock
\APACrefYear{2016}.
\newblock
\APACrefbtitle {{Hidden Markov Models for Time Series: An Introduction Using
  R}} {{Hidden Markov Models for Time Series: An Introduction Using R}}\
  (\PrintOrdinal{2}\ \BEd).
\newblock
\APACaddressPublisher{}{CRC press}.
\PrintBackRefs{\CurrentBib}

\end{thebibliography}

\end{document}